\documentclass{article}

\usepackage{PRIMEarxiv}

\usepackage{tikz}
\usetikzlibrary{shapes.geometric, arrows}
\usepackage{footmisc}

\usepackage{scrextend}

\usepackage{lscape}
\usepackage{pdflscape}
\usepackage[english]{babel}
\usepackage[utf8]{inputenc}
\usepackage{csquotes}
\usepackage{comment}
\usepackage[onehalfspacing]{setspace}
\usepackage{graphicx}
\usepackage{floatflt}
\usepackage{float}
\usepackage{setspace}
\usepackage{upgreek}
\usepackage{mathtools}
\usepackage{amssymb}
\usepackage{textcomp}
\usepackage{tabularx}
\usepackage{listings}
\usepackage[hidelinks]{hyperref}
\usepackage{abstract}
\usepackage{amsmath}
\usepackage{pdfpages}
\usepackage{acronym}
\usepackage{epigraph}
\usepackage[acronym, toc]{glossaries}
\usepackage{subfig}
\usepackage{listings}
\usepackage{xcolor}
\usepackage{listingsutf8}
\usepackage[T1]{fontenc}
\usepackage{multirow}

\makeglossaries

\usepackage{microtype}

\usepackage{rotating}

\definecolor{mygreen}{rgb}{0,0.6,0}
\definecolor{mygray}{rgb}{0.5,0.5,0.5}
\definecolor{mymauve}{rgb}{0.58,0,0.82}
\usepackage{orcidlink}
\usepackage[utf8]{inputenc} 
\usepackage[T1]{fontenc}    
\usepackage{hyperref}       
\usepackage{url}            
\usepackage{booktabs}       
\usepackage{amsfonts}       
\usepackage{nicefrac}       
\usepackage{microtype}      
\usepackage{lipsum}
\usepackage{fancyhdr}       
\usepackage{graphicx}       
\graphicspath{{media/}}     

\title{Deep Learning-Based Hypoglycemia Classification Across Multiple Prediction Horizons

}

\author{
  Beyza Cinar \orcidlink{0009-0006-3617-0134} \\ 
  Helmut Schmidt University, Germany\\
  \texttt{cinarb@hsu-hh.de} \\ 
   \And
  Jennifer Daniel Onwuchekwa \orcidlink{0000-0002-7721-3202} \\ 
  \texttt{j.danielonwuchekwa@hotmail.com} \\
   \And
   Maria Maleshkova \orcidlink{0000-0003-3458-4748} \\
   Helmut Schmidt University, Germany\\
   \texttt{maleshkm@hsu-hh.de}
}

\begin{document}
\maketitle

\begin{abstract}
Type 1 diabetes (T1D) management can be significantly enhanced through the use of predictive machine learning (ML) algorithms, which can mitigate the risk of adverse events like hypoglycemia. Hypoglycemia, characterized by blood glucose levels below 70 mg/dL, is a life-threatening condition typically caused by excessive insulin administration, missed meals, or physical activity. Its asymptomatic nature impedes timely intervention, making ML models crucial for early detection. This study integrates short- (up to 2h) and long-term (up to 24h) prediction horizons (PHs) within a single classification model to enhance decision support. The predicted times are 5-15 min, 15-30 min, 30 min-1h, 1-2h, 2-4h, 4-8h, 8-12h, and 12-24h before hypoglycemia. In addition, a simplified model classifying up to 4h before hypoglycemia is compared. We trained ResNet and LSTM models on glucose levels, insulin doses, and acceleration data. The results demonstrate the superiority of the LSTM models when classifying nine classes. In particular, subject-specific models yielded better performance but achieved high recall only for classes 0, 1, and 2 with 98\%, 72\%, and 50\%, respectively. A population-based six-class model improved the results with at least 60\% of events detected. In contrast, longer PHs remain challenging with the current approach and may be considered with different models.
\end{abstract}


\section{Introduction}  
Hypoglycemia is a life-threatening condition of low blood glucose levels below 70 mg/dL, affecting patients with type 1 diabetes (T1D). T1D is an incurable autoimmune disease with destroyed insulin production, developing in childhood and significantly impacting daily life. Since T1D patients cannot produce enough insulin, external injections are required to prevent hyperglycemia and glucose levels above 180 mg/dL \cite{ADA2010}. However, a major side effect is hypoglycemia, which can cause dizziness, coma, or death. Other contributing factors include low food intake and intense activity, which can affect insulin sensitivity \cite{Piersanti2023, Glumcevic2023}. Hypoglycemia can be prevented with timely glucose intake, but detection is challenging, especially during asymptomatic events like nocturnal hypoglycemia \cite{Bhimireddy2020}. Continuous glucose monitoring (CGM) devices, which measure glucose in interstitial fluid \cite{Swapna2021}, can provide input for predictive machine learning (ML) models to enable early intervention and personalized diabetes management \cite{Piersanti2023}. Integrating data from insulin pumps and activity trackers can further improve prediction accuracy \cite{Felizardo2021Review, Diouri2021}.
\\
Most studies employ binary classification detecting single PHs for separate use cases ranging from short-term prediction horizons (PHs) up to 2 hours until longer PHs up to 24h. However, no study has combined short- and long-term PHs within the same model for a multipurpose system. A model predicting hypoglycemia risk from 24h to 15 min before onset could enable immediate preventive actions and adjustments to daily activities, meal intake, and insulin dosages. This study integrates multiple PHs into a deep learning (DL) model to classify the time to hypoglycemic events in patients with T1D. It incorporates glucose levels, basal, and bolus insulin doses, and physical activity and compares CNN and RNN architectures.
\\
This study aims to decrease the risk of hypoglycemic events by alerting patients with T1D beforehand. In particular, the following contributions are made: 1) It integrates short- and long-term PHs into one classification model. 2) It investigates the performance of deep learning models while comparing ResNet, LSTM, and a hybrid model. 3) It compares between population-based and subject-specific models. 
\\
Finally, the structure is as follows. Section \ref{chap:StateOfArt} explores the state of the art in hypoglycemia estimation. The methodology is presented in section \ref{chap:Methods} by describing the utilized data, its preprocessing steps, and the model architectures. The results are presented in section \ref{chap:Results} while section \ref{chap:Discussion} discusses the findings, and section \ref{chap:Conlusion} concludes the main contributions of this work and highlights possible future work.

\section{State of the Art}
\label{chap:StateOfArt}

Hypoglycemia prediction can be approached with glucose forecasting and threshold detection using regression models \cite{Georga2013}. Alternatively, classification models identify hypoglycemia onset based on labeled data, often with less variability than regression models \cite{Zhang2024}. Existing classification approaches mainly focus on specific use cases, such as continuous, postprandial, nocturnal hypoglycemia, or 24-hour risk assessment.
For instance, Dave et al. used a random forest (RF) model for separate binary classifications at 15 min, 30 min, 45 min, and 1h PHs. They report sensitivities and specificities of at least 95\% \cite{Dave2020}. Vehì et al. employed support vector machines (SVMs) to classify hypoglycemia risk 4h in advance, achieving sensitivity and specificity of 69\% and 80\% for level 1 (below 70 mg/dL) and 75\% and 81\% for level 2 (below 54 mg/dL) hypoglycemia, respectively \cite{Vehi2019}. In addition, nocturnal hypoglycemia prediction 6h before onset is classified using ANNs, which achieve sensitivity of 44\% and specificity of 86\% \cite{Vehi2019}. Similarly, Bertachi et al. and Parcerisas et al. identify nocturnal hypoglycemia using SVMs, achieving sensitivity and specificity around 80\% \cite{Bertachi2020, Parcerisas2022}. Daily hypoglycemia risk has been addressed using decision trees and hybrid models, with sensitivity and specificity values of 87\% and 76\%, and 45\% and 89\%, respectively \cite{Piersanti2023, Felizardo2023}.
\\
Overall, sensitivity is often lower than specificity, and most studies rely on conventional ML models due to limited dataset sizes, typically involving 10–50 subjects. It can be further seen that the model architecture and chosen dataset are very influential. ML models require extensive feature selection, whereas DL models, such as LSTM and CNNs, can autonomously extract temporal patterns from glucose time-series data. 
Contrariwise, DL models such as LSTM and CNN are more often used for forecasting, such as in \cite{MunozOrganero2020, Seo2021, Jaloli2022}. In general, a classification of around 80\% of hypoglycemic events is noticed, while increasing PHs decrease performance. 

\section{Methodology}
\label{chap:Methods}

This study integrates multiple PHs into a single model to predict the time to hypoglycemia onset, enabling the alert of a risk as long as it continues. Utilizing multiple PHs can assist in planning daytime activities, optimizing meal and insulin adjustments, and facilitating short-term prevention.
\\
This section presents the used dataset, applied preprocessing steps, and the assignment of classes. Then, the model architectures are explored\footnote{The code can be found in \href{https://github.com/Mirai22/Hypoglycemia_Detection_MA}{https://github.com/Mirai22/Hypoglycemia\_Detection\_MA}}, and finally, employed metrics are introduced.

\subsection{Dataset: OhioT1DM}
\label{sub:Dataset}

The OhioT1DM dataset was used for its inclusion of glucose, insulin, and exercise data. Data were collected in free-living conditions for 8 weeks from six patients in 2018 and another six in 2020. A CGM device estimated glucose values every five minutes, while an insulin pump recorded bolus and basal insulin doses. Moreover, the 2018 cohort wore Basis Peak fitness bands, which collected data at 5-minute intervals, whereas the 2020 cohort used Empatica Embrace devices, collecting data at 1-minute intervals. Heart rate was recorded only by the Basis Peak, making it unusable as a feature. Additionally, the Basis Peak estimated step counts, while the Empatica Embrace measured the magnitude of acceleration \cite{OhioT1DM}.
Table \ref{tab:Dataset2018} and \ref{tab:Dataset2020} summarize the total number of samples and hypoglycemic data points ($\leq70$ mg/dL) for the 2018 and 2020 cohorts, respectively. Notable variations can be seen among subjects, potentially leading to a bias in the model. Specifically, subjects 540, 567, and 575 experienced more hypoglycemic states, and subjects 544, 584, and 588 significantly less. Lastly, the total duration of missing glucose data are summarized per patient.
\begin{table}[!ht]
\caption[Number of Samples in the OhioT1DM Dataset: 2018]{Number of Samples in the OhioT1DM Dataset: 2018 \label{tab:Dataset2018}}
\setlength\extrarowheight{-2pt}
    \begin{tabular}{| p{6.5cm}  |  p{1.1cm}  | p{1.1cm}  | p{1.1cm}  | p{1.1cm}  |p{1.1cm}  |p{1.1cm}  |  }
    \hline

    \textbf{Subject ID} & 559 & 563 & 570 & 575 & 588 & 591 \\ \hline    
    \textbf{Number of Hypoglycemic events} & 518 & 329 & 227 & 1173 & 136 & 570 \\  \hline
    \textbf{Total time of missing glucose values in days} & 5.7 & 3.8 & 2.6 & 4.7 & 1.9 & 6.9 \\  \hline
    \textbf{Number of Samples}  & 12792 & 14365 & 13500 & 13283 & 15295 & 13037 \\ \hline
\end{tabular}
\end{table}

\vspace{-2em}

\begin{table}[!ht]
\caption[Number of Samples in the OhioT1DM Dataset: 2020]{Number of Samples in the OhioT1DM Dataset: 2020 \label{tab:Dataset2020}}
\setlength\extrarowheight{-2pt}
    \begin{tabular}{| p{6.5cm} |  p{1.1cm}  | p{1.1cm}  | p{1.1cm}  | p{1.1cm}  |p{1.1cm}  |p{1.1cm}  |  }
    \hline
    \textbf{Subject ID} & 540 & 544 & 552 & 567 & 584 & 596 \\ \hline    
    \textbf{Number of Hypoglycemic events} & 986  & 188 & 408 & 925 & 137 & 300 \\ \hline
    \textbf{Total time of missing glucose values in days} & 4.6 & 8.6 & 12.5 & 11 & 5 & 10.5 \\  \hline
    \textbf{Number of Samples}  & 14843 & 13339 & 11444 & 13247 & 14815 & 13620 \\ \hline
\end{tabular}
\end{table}

\subsubsection{Preprocessing}
\label{sub:Preprocessing}

Wearable data can include noise, outliers, and missing values, which require preprocessing before the training on ML models. In particular, different frequencies or different data storage methods can cause gaps. Usually, missing values are imputed with linear interpolation and extrapolation. In addition, larger gaps can be removed entirely \cite{Nemat2023, Bertachi2020, Jaloli2022}. Tables \ref{tab:Dataset2018} and \ref{tab:Dataset2020} show that each patient has multiple days of missing values over the study duration.
\vspace{0.5em}

\textit{\textbf{Preprocessing:}} Before addressing missing values, the data were preprocessed, and individual parameters were merged into a single dataset. Glucose levels, insulin doses, and acceleration data were rounded to 5-minute intervals and resampled to the same frequency. For basal insulin, continuous and temporally applied doses were initially resampled to 5-minute intervals using forward filling with the previous value. These two sources were merged, replacing overlapping temporal basal dosage entries with their corresponding values. Due to its periodic administration, missing basal insulin data were imputed using forward and backward filling. For bolus insulin, the start and end times of infusion were recorded. The reported dosage was applied to recorded start and end times of infusion at these intervals. Since not reported times indicate no bolus was infused, missing values were filled with zero.
To include physical activity, the step count measured for the 2018 cohort was first converted into the magnitude of acceleration. This approximation was calculated using step count and time data, applying the following formulas \cite{Velocity, Acceleration}:
\begin{equation*}
distance(x) = step\_count(x) * 0.75
\end{equation*}
\begin{equation}
velocity(x) = distance(x) / time
\end{equation}
\begin{equation*}
acceleration(x1) = (velocity(x1) - velocity(x0)) / time \_intervall. 
\end{equation*}
Acceleration is the change in velocity divided by the change in time. Since the steps were measured every 5 minutes, the time for each distance and the time difference between velocity changes in 5 minutes \cite{Velocity, Acceleration}. Based on the literature, the step is converted to meters by multiplying by 0.75, which is used as the distance unit. Lastly, the acceleration data of both cohorts were scaled to values between 0 and 1 to have uniform data. \vspace{1em}

\textit{\textbf{Annotation of Classes:}} To define hypoglycemia, each sample less and equal to 70 mg/dL was assigned class 0. Then, 5-15 min before hypoglycemia was annotated with class 1, more than 15-30 min before was classified as class 2, more than 30-60 min before as class 3, more than 1-2h before as class 4, more than 2-4h before as class 5, more than 4-8h before as class 6, more than 8-12h before as class 7, and more than 12-24h before as class 8. To ensure that no instance is overwritten with a new class, all instances were first assigned a value of $-1$ as the class. Then, the hypoglycemia threshold was applied over all samples to define the hypoglycemic state. Thereafter, the time condition was only used for instances assigned to class $-1$. \vspace{1em}

\textit{\textbf{Handling Missing Data:}} Linear interpolation was employed to fill missing data points between two known values of the same parameter \cite{Noor2013}. 
Most missing values appeared for glucose and then for acceleration. Only missing values less than 120 min were interpolated since data could be overestimated. The remaining missing values for glucose were removed so that each patient had a collection of multiple data-frames without any time gaps. Contrariwise, missing acceleration values were indicated as missing values by replacing them with $-1$. \vspace{1em}

\subsubsection{Time Series Generation}
\label{sub:PreDataAnalysis}

Time series sequences were created for the train and test datasets with a predefined input sequence length (ISL). For labeling the series, the class of the last sample was used. Experiments in section \ref{sub:Model} revealed that for classifying up to 24h, ISLs of 6 and 8h yield better performance. 
The distribution of samples per class after the generation of time series sequences with a length of 8h can be seen in Table \ref{tab:ClassSamplesGap}. Creating sequences leads to lost instances due to unusable data-frames caused by gap removal. Longer sequences either have inefficient computation time or result in even fewer samples. 
\vspace{-2em}
\begin{table}[!ht]
\caption[Samples per Class for Preprocessed Time Series Data of 8h]{Samples per Class for Preprocessed Time Series Data of 8h}
\label{tab:ClassSamplesGap}
\setlength\extrarowheight{-3pt}
    \begin{tabular}{| p{0.8cm} |  p{1.2cm}  |  p{0.8cm}  | p{0.8cm}  | p{0.8cm}|  p{0.8cm}  | p{0.8cm}  | p{0.8cm}  | p{0.8cm}|  p{0.8cm}  | p{0.8cm}  | p{0.8cm}  |}
    \hline

     &  & 
    \multicolumn{10}{ c |} {\textbf{Class}}\\ 
    \cline{3-12} \textbf{ID} & \textbf{Total} &  \textbf{0} & \textbf{1} & \textbf{2} & \textbf{3}  & \textbf{4} & \textbf{5}  & \textbf{6} & \textbf{7} & \textbf{8} & \textbf{9}  \\  \hline 

    \textbf{540} &13722& 894 &291 & 277 & 523 &933 &1521 &2418&1640&3086&2139\\ \hline
    
    \textbf{544} &7366& 184&78&69&110&192&398&842&861&2073&2559 \\ \hline

    \textbf{552} &9168 &418 & 130 & 125 & 237 & 466 & 908 &1520 &1176 &2154 & 2034\\ \hline

    \textbf{559} &11450& 441&152&138&268&522&1037&1950&1589&3066&2287  \\ \hline
    
    \textbf{563} & 10613&327&124&118&217&415&685&1283&1114&2577&3753 \\ \hline

    \textbf{567} &11747&932 &137&120&226&429&830&1495&1242&2698 &3638 \\ \hline
    
    \textbf{570} &7050&227 &87&77&135&238&396&720&686&1804& 2680\\ \hline
    
    \textbf{575} &13738& 1158&261&236&437&834&1488&2538&1777&3140&1869 \\ \hline
    
    \textbf{584} &4921& 130&70&64&126&236&404&589&528&1370& 1404 \\ \hline

    \textbf{588} &6078&138 &77&68&132&238&432&757&645&1448&2143 \\ \hline
    
    \textbf{591} &12919& 600&214&197&382&731&1263&2110&1807&3419& 2196 \\ \hline

    \textbf{596} &10384&259 &136&122&218&425&788&1412&1273&2850& 2901\\ \hline

\end{tabular}
\end{table}

\subsection{Model Architectures}
\label{sub:Model}

This work developed a population-based DL model to classify the time to the onset of hypoglycemia. The main model was decided by validating on a randomly chosen subject (subject 552), used as the test data. Data from the remaining subjects were split into training and validation, while the validation data were the last 20\% of the individual training data. Then, data were shuffled. 
\\
Approached models included LSTM, BiLSTM, ResNet, and 1DCNN. For all models, early stopping, best weight saving and restoring, and learning rate reduction were implemented as callback functions based on the validation loss. The patience for learning rate reduction was set to 3 for all models, while early stopping was set to 5 for RNN and 10 for CNN models. Each model was trained using the Adam optimizer and the sparse categorical cross-entropy loss function. Simple LSTM and ResNet models were further tuned, with both architectures tested across various ISLs of 12, 8, 6, and 4 hours. A seed value of 42 was used to ensure reproducibility. Additionally, class weights were applied to address the class imbalance, with the following function, in which "class\_occurrence" counts the number of samples of the class $x$, "total" counts the number of all available samples, while "number\_of\_classes" refers to the set of classes~\cite{TensorWeightClass}: 
\begin{equation}
    weight_x = (1/class\_occurrence) * (total/number\_of\_classes)
\end{equation}

\subsubsection*{LSTM vs BiLSTM Model}

\begin{figure}[!ht]
	\centering
	\includegraphics[width=0.7\textwidth]{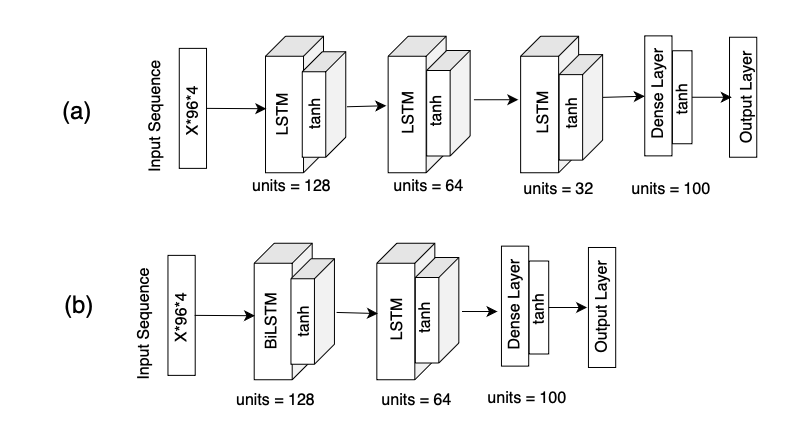}
    \caption[Architecture of the Applied RNN Models]{
    Architecture of the Applied RNN Models \begin{small} a) LSTM, b) BiLSTM \end{small}}
    \label{fig:ModelArchiRNN}
\end{figure}

Three LSTM models including one layer (128 units), two layers (128 and 64 units) and three layers (128, 64, and 32 units) were tested, with the three layered performing best. More layers increased computation time significantly, while dropout layers and global average 1D pooling decreased performance. It was trained with a hyperbolic tangent (tanh) activation function after comparing it with the rectified linear unit (ReLu), exponential linear unit (ELU), and swish. Among the tested batch sizes of 32, 64, and 128, batch size 64 showed better performance. An improvement was seen when adding a dense layer of 100 units following the last LSTM layer. For all reported experiments, an ISL of 12h was utilized. After testing different ISLs, 8h and 12h sequences produced similar results. However, the shorter computation time of the 8h sequence was advantageous. Finally, the best LSTM model was compared to a BiLSTM model. The architecture of compared RNN models can be seen in Fig. \ref{fig:ModelArchiRNN}. From Table \ref{tab:TestLSTM}, it is noticed that LSTM is superior with a shorter computation time and better recall for the first classes and class 5 \ref{tab:TestLSTM}.
\vspace{-2em}
 \vspace{-2em}

\begin{table}[!ht]
\caption[Comparison of LSTM and BiLSTM models]{Comparison of the LSTM and BiLSTM Models Trained With a Batch Size of 64 Across 9 Classes}{\label{tab:TestLSTM}}
\setlength\extrarowheight{-3pt}
    \begin{tabular}{| p{2cm} |  p{2.2cm}  | p{2.2cm}  | p{2.2cm}  | p{2.1cm}  |p{2.5cm} |  }
    \hline
    \textbf{Model} & \textbf{Input- Length} &  \textbf{Accuracy (\%)} & \textbf{Precision (\%)} & \textbf{Recall (\%)} & 
    \textbf{F1-Measure (\%)}  \\  \hline    
    LSTM & 12h & 30 & 33 & 42  & 35\\  \hline
    LSTM & 8h & 30& 33 & 42&35 \\  \hline
    BiLSTM & 12h & 31 &   32&41 & 33 \\  \hline
    BiLSTM & 8h & 32 & 33  & 42&  35\\  \hline
\end{tabular}
\end{table}

\subsubsection*{1DCNN vs ResNet Model}

\begin{figure}[!ht]
	\centering
	\includegraphics[width=1\textwidth]{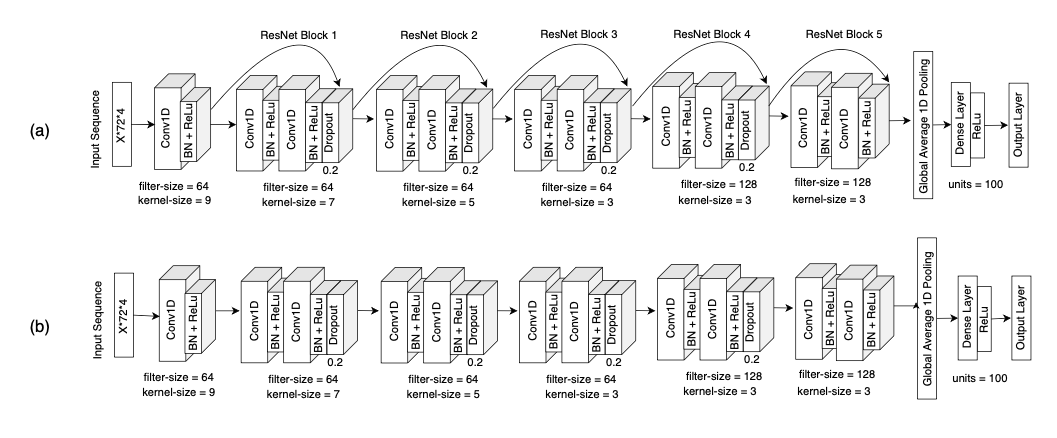}
    \caption[Architecture of the Applied CNN Models]{
    Architecture of the Applied CNN Models \begin{small} a) ResNet, b) 1DCNN \end{small}}
    \label{fig:ModelArchiCNN}
\end{figure}

After testing various kernel sizes and layer configurations, the ResNet model with five residual blocks, each containing two 1DCNN layers, outperformed the 1DCNN model. Each layer had the same kernel size, as shown in Fig. \ref{fig:ModelArchiCNN}. Before the first residual block, a 1DCNN layer with a kernel size of 9 was applied without a shortcut connection. All 1DCNN layers had 64 filters.
Each convolutional layer was followed by batch normalization (BN) and the Rectified Linear Unit (ReLU) activation function, outperforming ELU, Swish, and tanh activations. A dense layer with 100 units after the global 1D average pooling improved performance, whereas using a 1D max-pooling layer led to poorer results. Among batch sizes of 32, 64, 128, and 256, a batch size of 128 was selected. The performance also improved by adding dropout layers after each residual block, except the last one.
Further testing was conducted with three-block models using kernel sizes of 7, 5, and 3, respectively. The three-block model was tested with a filter size of 128 for the final residual block, while the five-block model used 128 filters for the fourth and fifth residual blocks. Models with three stacked convolutional layers performed worse. The best-performing configuration used filter sizes of 64, 64, 64, 128, and 128, with kernel sizes of 7, 5, 3, 3, and 3 for blocks 1 through 5, respectively.
Finally, different ISLs were evaluated, with a 6-hour ISL yielding the best results, as presented in Table \ref{tab:TestCNN}.
Thereafter, the same architecture without the shortcut connections was tested as a simple 1DCNN model, of which the results can be seen in Table \ref{tab:TestCNN}. The ResNet model has a slightly better classification performance. Thus, it was chosen as superior. Further, the metrics of the best LSTM and the best ResNet model do not vary much.  
\begin{table}[!ht]
\caption[Comparison of ResNet and 1DCNN models]{Comparison of the ResNet and 1DCNN Models Trained with a Batch Size of 128 Across 9 Classes}{\label{tab:TestCNN}}
\setlength\extrarowheight{-3pt}
    \begin{tabular}{| p{2cm} |  p{2.2cm}  | p{2.2cm}  | p{2.2cm}  | p{2.1cm}  |p{2.2cm} |  }
    \hline
    \textbf{Model} & \textbf{Input- Length} &  \textbf{Accuracy (\%)} & \textbf{Precision (\%)} & \textbf{Recall (\%)} & 
    \textbf{F1-Measure (\%)}  \\  \hline    
    
    ResNet & 12h & 29 & 32 & 36  & 33\\  \hline
    ResNet & 6h & 28 &33& 42 & 34 \\  \hline
    1DCNN & 12h & 27 & 27  & 33 & 28 \\  \hline
    1DCNN & 6h & 28 & 30 & 37 & 31 \\  \hline
\end{tabular}
\end{table}

\subsubsection*{Hybrid Model} 
\begin{figure}[!ht]
	\centering
	\includegraphics[width=1\textwidth]{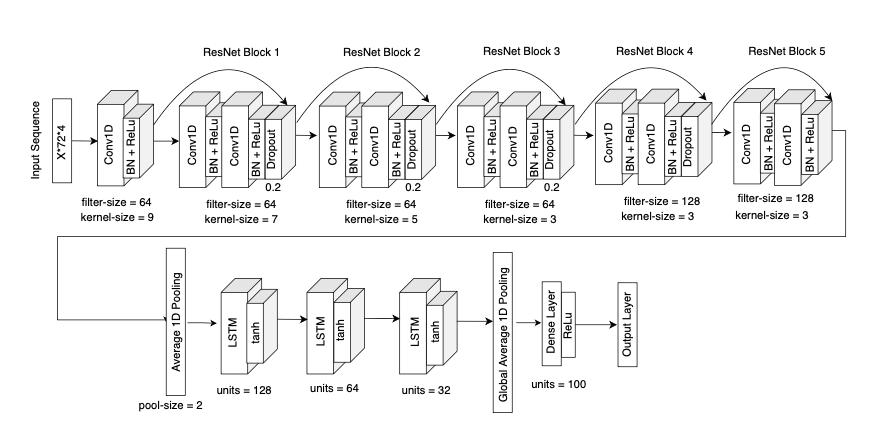}
    \caption[Architecture of the Applied Hybrid Model]{Architecture of the Applied Hybrid Model: ResNet + LSTM} 
    \label{fig:ModelArchiHyb}
\end{figure}

The hybrid model depicted in Fig. \ref{fig:ModelArchiHyb} is a stacked model consisting of the ResNet model built by one 1DCNN layer followed by five residual blocks and three LSTM layers. This model architecture was not tested prior on subject 552.

\subsection{Metrics Used for the Model Evaluation}
\label{sub:Metrics}

Multiclass classification systems require evaluation using metrics like accuracy, precision, recall, and F1-measure (F1-M). Additionally, confusion matrices are analyzed to provide insight into true positives (TP), false positives (FP), false negatives (FN), and true negatives (TN) for each class. Precision measures the proportion of TP among all instances classified as positive, while recall indicates the proportion of correctly identified samples within a given class. The F1-score, a harmonic mean of precision and recall, balances both metrics. Accuracy reflects the proportion of correctly classified samples across the entire dataset. However, it is biased with imbalanced datasets. This study reports macro-averaged metrics to address this imbalance, ensuring that each class is weighted equally \cite{Grandini2020MetricsFM}.

\section{Results}
\label{chap:Results}

The three chosen models were trained with a leave-one-out cross-validation with 12 folds, with each subject being a test fold once. The ResNet and hybrid models were trained with an ISL of 6h and a batch size of 128, while the LSTM model was trained with an ISL of 8h and a batch size of 64. For all models, early stopping with a patience of 10 was applied. Moreover, transfer learning was employed in which the model was further trained with a batch size of 16 with 50\% of the test subject's data. 20\% of the data were used for validation and 30\% for testing, with a patience for early stopping set to 20 epochs. In addition, the population-based models were tested with the same reduced test data to enable a fair comparison.
Lastly, the population- and individual-based models were trained with only 6 classes classifying up to 4h before hypoglycemia, using half of the prior used batch size and an ISL of 4h. 
\\
This section reports the population-based models utilizing nine classes in subsection \ref{sub:Pop9class} and compares them to person-specific models in subsection \ref{sub:9compare}. Finally, the models with reduced classes are presented in subsections \ref{sub:Pop6class} and~\ref{sub:6compare}.

\subsection{Population-Based Models Using 9 Classes}
\label{sub:Pop9class}

First, up to 24h before hypoglycemia was classified. Table \ref{tab:Comp9class} presents each model's population-based macro average (Avg) performances for each class. It can be seen that the LSTM model is superior, with a better classification of classes 0-2, 4, and 8. Classes 3, 5, and 7 have slightly increased metrics with the hybrid approach, while the ResNet model only produces better results for class 6. The macro averages for each subject reveal great variations between subjects, while the LSTM model obtains better results on average, especially for the recall and the F1-measure. On average, a worse precision than recall is obtained. Overall, the best performance for all metrics is achieved for subject 567. Nevertheless, considering all maximum values, the model has a poor classification ability and cannot even achieve a recall of 50\%, whereas the best F1-measure is only 39\%. 
\begin{table}
\centering
\caption[Population-Based LSTM Results for Each Subject Using 9 Classes]{Population-Based LSTM Results for Each Subject Using 9 Classes \label{tab:LSTMPop9}} 
\setlength\extrarowheight{-2pt}
\resizebox{!}{7cm}
{\begin{tabular}
{|p{1.5cm}|p{1.5cm}|p{1.2cm}|p{1.2cm}|p{1.2cm}|p{1.2cm}|p{1.2cm}|p{1.2cm}|p{1.2cm}|p{1.2cm}|p{1.2cm}|}
\cline{1-11} \textbf{Subject} & \textbf{Metric} & 
\multicolumn{9}{ l |} {\textbf{Class}}\\ 
\cline{3-11} & & 0 & 1  & 2 & 3 & 4 & 5 & 6 & 7  & 8  \\ \hline
    
\multicolumn{1}{|l|}{\multirow{3}{*}{}}  & Precision & \textbf{1.00} &  0.42   & 0.16 & 0.11 & 0.12 & 0.15 & 0.29 & 0.19 & 0.35 \\ 

\cline{2-11} 
\multicolumn{1}{|l|}{\textbf{540}}   & Recall & 0.98  & 0.70 & 0.54 & 0.18 & 0.26 & 0.15 &  0.20& 0.18 & 0.20  \\ 

\cline{2-11} 
\multicolumn{1}{|l|}{} & F1-M & 0.99 & 0.53 & 0.25 & 0.14&  0.16& 0.15 & 0.23   & 0.19 & 0.26 \\  \hline\hline

\multicolumn{1}{|l|}{\multirow{6}{*}{}}  & Precision  &  0.97  & 0.35  & 0.16 & 0.09 & 0.00 & 0.00 & 0.20 & 0.16 & 0.48 \\ 

\cline{2-11}  
\multicolumn{1}{|l|}{\textbf{544}}   & Recall& 0.80 & 0.47 & \textbf{0.71}  & 0.23 & 0.00 & 0.00 & 0.24 & 0.05 & \textbf{0.63}  \\ 

\cline{2-11} 
\multicolumn{1}{|l|}{} & F1-M&0.88 & 0.40 & 0.26 & 0.13 &  0.00& 0.00 & 0.22 & 0.07   & \textbf{0.54}\\  \hline\hline

\multicolumn{1}{|l|}{\multirow{6}{*}{}}  &Precision  & 0.99   & 0.47  & 0.18 & 0.13 & 0.15 & 0.18 & 0.28 & 0.10 & 0.43 \\ 

\cline{2-11}  
\multicolumn{1}{|l|}{\textbf{552}}   & Recall& 0.97 & 0.66 & 0.62     & 0.36 & 0.22 & \textbf{0.40}  & 0.22 & 0.06 & 0.23  \\ 

\cline{2-11} 
\multicolumn{1}{|l|}{} & F1-M& 0.98& 0.55 &0.28  & 0.19 & 0.18 & \textbf{0.25} & 0.25& 0.07   & 0.30 \\  \hline\hline

\multicolumn{1}{|l|}{\multirow{6}{*}{}}  &Precision  & \textbf{1.00} & 0.40    &0.16  & 0.13 & 0.14 & \textbf{0.19} & 0.26 & 0.16 & 0.42 \\ 

\cline{2-11}  
\multicolumn{1}{|l|}{\textbf{559}}   &Recall& 0.97 &0.70  &0.61 & 0.37 & 0.32 & 0.17 & 0.15 & 0.16 & 0.32  \\ 

\cline{2-11}  
\multicolumn{1}{|l|}{} & F1-M & 0.99& 0.51 & 0.25 & 0.20 & 0.19 & 0.18 & 0.19 & 0.16 & 0.36 \\  \hline\hline

\multicolumn{1}{|l|}{\multirow{6}{*}{}}  & Precision &  \textbf{1.00}  & 0.40  & 0.16 & 0.13 & 0.15 & 0.15 & 0.21 & 0.25 & 0.48 \\ 

\cline{2-11}  
\multicolumn{1}{|l|}{\textbf{563}}   & Recall& 0.99 & 0.75 & 0.51 &  0.29&  0.32& 0.13 & 0.19 & 0.17 & 0.35  \\ 

\cline{2-11} 
\multicolumn{1}{|l|}{} & F1-M & 0.99& 0.52 & 0.24 & 0.18 & 0.20 & 0.14 & 0.20  & 0.20 & 0.40 \\  \hline\hline

\multicolumn{1}{|l|}{\multirow{6}{*}{}}  & Precision  & \textbf{1.00} & \textbf{0.58}   & \textbf{0.25} & \textbf{0.21} & 0.15 & 0.15 & \textbf{0.33} & 0.22 &0.48  \\ 

\cline{2-11} 
\multicolumn{1}{|l|}{\textbf{567}}   & Recall& \textbf{1.00} & \textbf{0.77} & 0.69 & 0.44 & 0.41 & 0.24 & 0.12 &0.25  &0.32   \\ 

\cline{2-11} 
\multicolumn{1}{|l|}{} & F1-M & \textbf{1.00}& \textbf{0.66} & \textbf{0.37} & \textbf{0.29} & 0.22 & 0.18 & 0.18   & \textbf{0.23} & 0.39 \\  \hline\hline

\multicolumn{1}{|l|}{\multirow{6}{*}{}}  & Precision  &   \textbf{1.00} & 0.36  & 0.16 &  0.18& \textbf{0.24} & 0.12 & 0.22 & 0.17 & 0.40  \\ 

\cline{2-11} 
\multicolumn{1}{|l|}{\textbf{570}}   & Recall& 0.80 & 0.60 & 0.33 & 0.30 & 0.20 & 0.06 & \textbf{0.57} & 0.25 & 0.09  \\ 

\cline{2-11} 
\multicolumn{1}{|l|}{} & F1-M & 0.89& 0.45 & 0.22 & 0.23 & 0.22 & 0.08 & \textbf{0.32} & 0.20 & 0.14 \\  \hline\hline

\multicolumn{1}{|l|}{\multirow{6}{*}{}}  & Precision    & 0.94 & 0.30  & 0.12 & 0.19 & 0.17 & 0.11 & 0.25& \textbf{0.26} & 0.32 \\ 

\cline{2-11} 
\multicolumn{1}{|l|}{\textbf{575}}   &Recall & \textbf{1.00} & 0.71 &  0.34 & 0.26 & 0.46 & 0.00 & 0.24 & 0.05 & 0.37  \\ 

\cline{2-11} 
\multicolumn{1}{|l|}{} & F1-M & 0.97 & 0.42 & 0.17 & 0.22 & \textbf{0.25} & 0.00   & 0.24 & 0.08 & 0.34 \\  \hline\hline

\multicolumn{1}{|l|}{\multirow{6}{*}{}}  & Precision  &  0.99  &0.41   & 0.11 & 0.08 & 0.08 & 0.13 & 0.12 & 0.09 & 0.38 \\ 

\cline{2-11} 
\multicolumn{1}{|l|}{\textbf{584}}   & Recall & \textbf{1.00} & 0.58 & 0.37 & 0.18 & 0.16 & 0.10 & 0.14 & 0.07 & 0.24  \\ 

\cline{2-11} 
\multicolumn{1}{|l|}{} & F1-M &0.99 & 0.48 & 0.17 & 0.11& 0.11 & 0.11 & 0.13  & 0.08 & 0.29 \\  \hline\hline

\multicolumn{1}{|l|}{\multirow{6}{*}{}}  & Precision  & 0.93   & 0.44  & 0.11 & 0.11 &0.13  & 0.15 & 0.20 & 0.12 & \textbf{0.55} \\ 

\cline{2-11} 
\multicolumn{1}{|l|}{\textbf{588}}   & Recall & \textbf{1.00} & 0.59 & 0.36  & 0.23 & \textbf{0.55} & 0.11 & 0.28 & 0.02 & 0.21    \\ 

\cline{2-11} 
\multicolumn{1}{|l|}{} & F1-M & 0.96&  0.50&  0.17& 0.15 & 0.21 &0.13  & 0.23   & 0.03 & 0.30 \\  \hline\hline

\multicolumn{1}{|l|}{\multirow{6}{*}{}}  & Precision & \textbf{1.00}  & 0.37  & 0.14 & 0.13 & 0.11 & 0.13 & 0.29 & 0.18 & 0.42 \\ 

\cline{2-11}  
\multicolumn{1}{|l|}{\textbf{591}}   & Recall & 0.80 & 0.50 & 0.44  & 0.40 & 0.08 & 0.21& 0.20 & 0.14 & 0.32  \\ 

\cline{2-11} 
\multicolumn{1}{|l|}{} & F1-M& 0.89& 0.43 & 0.21 & 0.19 & 0.09 & 0.16   & 0.23 & 0.15 & 0.37 \\  \hline\hline

\multicolumn{1}{|l|}{\multirow{6}{*}{}}  & Precision  & 0.99  & 0.44  & 0.19 & 0.09 & 0.11 & 0.11 & 0.13& 0.19 & 0.40 \\ 

\cline{2-11} 
\multicolumn{1}{|l|}{\textbf{596}}   & Recall & \textbf{1.00} & 0.75 & 0.46 & \textbf{0.49} & 0.11 & 0.14 & 0.01 & \textbf{0.26} & 0.33  \\ 

\cline{2-11} 
\multicolumn{1}{|l|}{} & F1-M & 0.99& 0.56 & 0.27 & 0.15 & 0.11 & 0.12 &0.02  & 0.22   & 0.36 \\  \hline

\end{tabular}}
\end{table}
\begin{table}
\centering
\caption[Performance of Each Class for Each Model Using 9 Classes]{Performance of Each Class for Each Model Using 9 Classes \label{tab:Comp9class}}
\setlength\extrarowheight{-2pt}
\resizebox{!}{2.5cm}
{\begin{tabular}{p{0.5cm}|p{1.5cm}|p{1.2cm}|p{1.2cm}|p{1.2cm}|p{1.2cm}|p{1.2cm}|p{1.2cm}|p{1.2cm}|p{1.2cm}|p{1.2cm}|p{1.2cm}|}
\cline{2-12}
    & \textbf{Metric} & \textbf{Avg} &
    \multicolumn{9}{ l |} {\textbf{Class}}\\ 
    \cline{4-12} & & & 0 & 1  & 2 & 3 & 4 & 5 & 6 & 7  & 8  \\ \hline

\cline{2-12} 
\multicolumn{1}{|l|}{\multirow{4}{*}{}} & Precision & 0.30 &0.92 &0.35  &0.15&0.13&0.13& 0.14&0.24&0.18 & 0.43    \\  

\cline{2-12}  
\multicolumn{1}{|l|}{\textbf{ResNet}} & Recall  &0.34  &0.93 &0.60  &0.36&0.19&0.13 &0.11& 0.27& 0.30  &0.21   \\ 

\cline{2-12}  
\multicolumn{1}{|l|}{}  & F1-M  & 0.30 &0.92 & 0.44 &0.21&0.15& 0.13&0.12 &0.25   &0.22 & 0.28 \\   
 \hline\hline

\multicolumn{1}{|l|}{\multirow{4}{*}{}} & Precision & 0.31 &0.98 & 0.39 &0.16&0.13&0.14 &0.15& 0.24& 0.18  &0.42  \\ 

\cline{2-12}  
\multicolumn{1}{|l|}{\textbf{LSTM}}    & Recall  & 0.39 &0.96 & 0.66 &0.49&0.31& 0.27&0.15& 0.20&0.14   &0.31 \\  

\cline{2-12}  
\multicolumn{1}{|l|}{} & F1-M  &0.33  &0.97 & 0.49 &0.24&0.18& 0.18&0.15& 0.22& 0.16  & 0.35 \\    
\hline\hline

\multicolumn{1}{|l|}{\multirow{4}{*}{}} & Precision & 0.30 &0.95 & 0.27 &0.16&0.14&0.13 &0.16&0.23 &0.22   &0.42  \\

\cline{2-12}  
\multicolumn{1}{|l|}{\textbf{Hybrid}}   & Recall & 0.35 &0.83 & 0.47 &0.39&0.30&0.25&0.21 &0.09   & 0.33 & 0.26  \\  

\cline{2-12}  
\multicolumn{1}{|l|}{} & F1-M  &  0.30&0.88 &0.34  &0.22&0.19&0.17 &0.18&0.13 &0.26   &0.32 \\     
\hline

\end{tabular}}
\end{table}
\\In the following, the performance of the single classes is investigated for a population-based (PB) LSTM model, which is identified as the best approach. Table \ref{tab:LSTMPop9} presents the precision, recall, and F1-measure of each class and each subject, with the best values of each metric among the subjects being highlighted. 
It is noticed that the performance of the first classes is increased compared to the latter classes, which are less distinguishable. In particular, the first class is detected correctly with a recall of at least 97\% while only three subjects achieve less. The precision behaves similarly with the worst F1-measure of 88\% in subject 544 and the best of 100\% in subject 567. Turning now to the performance of class 1, there is a decrease in the performance, along with more significant variation between subjects. In total, 6 subjects have a recall of at least 70\%, with a population recall of 66\%. Contrariwise, an average low precision of 39\% induces many false alarms. On average, more than 60\% of all hypoglycemic cases are detected up to 15 min before but with an F1-measure of 49\%. The best performance is seen in subject 567 with an F1-measure of 66\%, while the worst performance is in subject 544 with an F1-measure of 40\%.
More variations can be observed in class 2 with a population-based precision, recall, and F1-measure of 16\%, 49\%, and 24\%, respectively. The worst F1-measure is 17\%, while the best is 37\%. 
For the following classes, the performance decreases radically. For class 3, the best F1-measure is only 29\%, while the population-based mean F1-measure is 18\%, highlighting a significant difference between recall and precision. The worst metrics are 8\%, 18\%, and 11\% for precision, recall, and F1-measure, respectively, in subject 584. 
Likewise, class 4 obtains its best precision with only 24\%, the best recall with 55\%, and the best F1-measure with 25\%. The worst performance is seen in subject 544, with 0\% in all metrics. The population-based F1-measure is 18\%. 
Class 5 performs very poorly with a population precision, recall, and F1-measure of 15\%. Subject 552 is above the average with an F1-measure of 25\%.
Moving to class 6, the best individual values are 22\%, 57\%, and 32\%, and the population-based metrics are 24\%, 20\%, and 22\%, respectively, for precision, recall, and F1-measure. Besides, class 7 obtains its best F1-measure with 23\% while the population F1-measure is only 16\%.
Lastly, class 8 has an improved performance with an average F1-measure of of 35\%, a best individual F1-measure of 54\%, and a worst F1-measure of 14\%. 
\begin{figure}[!ht] 
	\centering
    \captionsetup{justification=centering}
    \subfloat[][]{\includegraphics[width=0.35\textwidth]{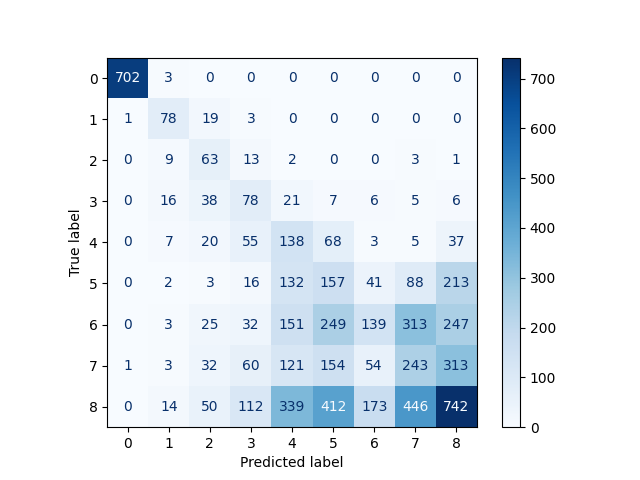}}
    \hspace{0.2cm}
    \captionsetup{justification=centering}
    \subfloat[][]{\includegraphics[width=0.35\textwidth]{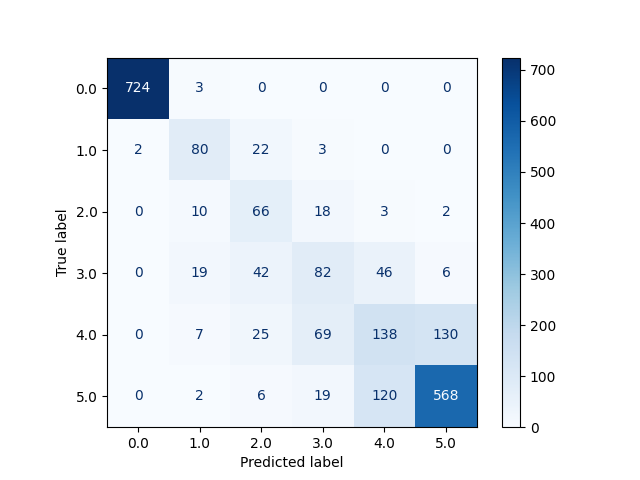}}
    \caption[Population-Based Confusion-Matrix Across 9 and 6 Classes]{Population-Based Confusion-Matrix Across 9 and 6 Classes\\ of Subject 567 Trained with the LSTM Model \label{fig:CMPop9}}
\end{figure}
Using a ResNet model, the distribution of best and worst metrics differ across subjects, with stronger variations between classes and subjects. Furthermore, the overall performance decreases with a hybrid model, but the distribution among subjects is similar to the LSTM model. 
\\
To conclude, despite having fewer samples, the first classes are better classified as to why the model can only identify the hypoglycemic event 0-30 min before with an acceptable recall but decreased precision. Since early stopping is applied, it is determined that subjects who trained for more epochs also achieved better classification results. 
\\
Finally, looking at the confusion matrix of the subjects with the best results in Fig. \ref{fig:CMPop9}, it can be seen that for the LSTM model, class 0 is only misclassified with class 1, while most of the samples of class 1 are classified correctly or as class 2. Furthermore, the model cannot distinguish samples belonging to class 3 from class 1 or 2 with high confidence, while more instances are classified correctly. Starting with class 4, the performance decreases, and low precision is noticed. In short, misclassifications are more often with the nearest neighbors.

\subsection{Subject-Specific and Population-Based Models Using 9 Classes}
\label{sub:9compare}

\begin{figure}[!ht]
	\centering
    \captionsetup{justification=centering}
    \subfloat[][]{\includegraphics[width=0.33\textwidth]{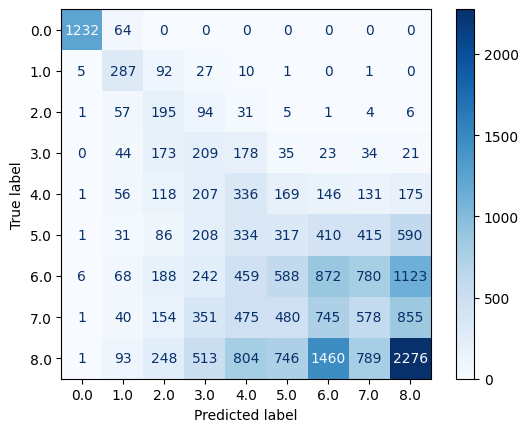}}
    \hspace{0.3cm}
    \subfloat[][]{\includegraphics[width=0.4\textwidth]{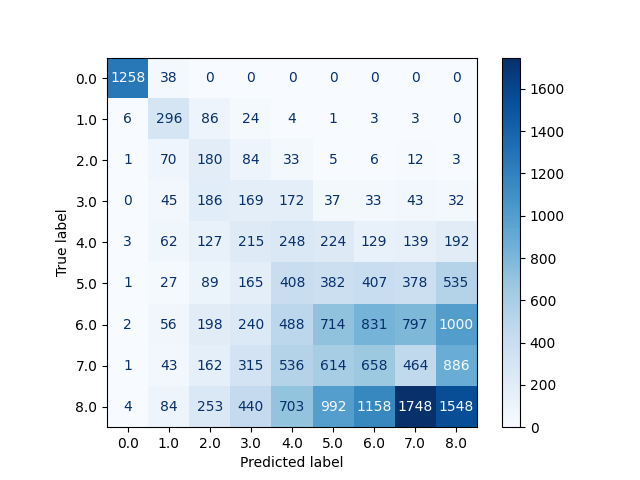}}
    \hspace{0.3cm}
    \subfloat[][]{\includegraphics[width=0.33\textwidth]{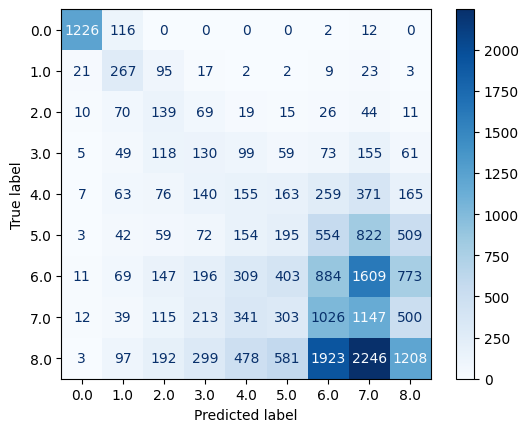}}
    \hspace{0.3cm}
    \subfloat[][]{\includegraphics[width=0.4\textwidth]{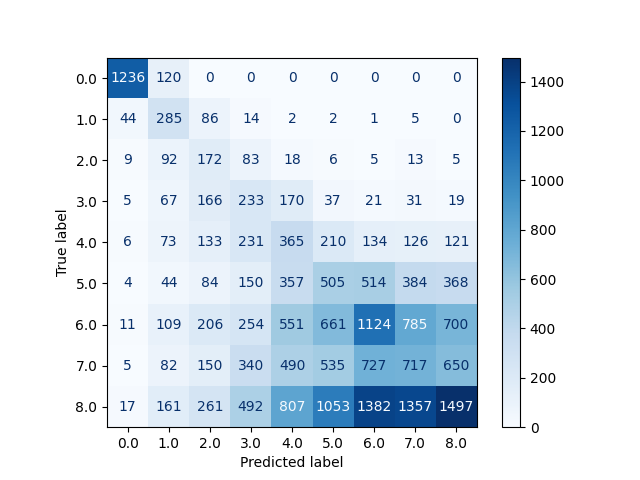}}
    \caption[Population-Based and Subject-Specific Confusion-Matrices Across 9 Classes]{Population-Based and Subject-Specific Confusion-Matrices Across 9 Classes \label{fig:CMAll9}}
	 (a) Population-Based LSTM Model (b) Subject-Specific LSTM Model (c) Population-Based ResNet Model (d) Subject-Specific ResNet Model 
\end{figure}

To address significant variations between subjects, subject-specific (SS) models were developed using transfer learning. Examining the population metrics of each model, as presented in Table \ref{tab:Comp9classlesstest}, it can be seen that the performance of both LSTM models is similar. However, the subject-specific model improves all metrics for classes 0, 1, and 5, while class 2 shows better recall but lower precision and F1-score. Class 3 remains unchanged across all metrics, whereas classes 4, 6, 7, and 8 perform better with the population-based model. Individualized training also improved the performance of the ResNet model, but the LSTM model remains superior. The hybrid model improves in accuracy and macro-average recall when trained on subject-specific data. In contrast, the population-based hybrid model yields results similar to the LSTM model.
No model achieves consistently high classification performance across all classes, indicating that the overall classification ability of long-term PHs remains insufficient. 
The macro averages for each subject reveal that the LSTM model remains the best-performing approach, with superior overall results in four subjects and the highest F1-scores across all subjects. Hence, the LSTM model is investigated in more detail. 
\\
Table \ref{tab:LSTMPop9less} presents the performance of single classes for each subject using nine classes with fewer data, and the results of the subject-specific model can be seen in Table \ref{tab:LSTMIndi9}. Comparing both methods, it is noticed that individualization improves the overall performance of most subjects. Notably, for class 0, the lowest values of the population-based models improve from 93\%, 76\%, and 86\%, to 97\%, 89\%, and 94\% for the subject-specific models for precision, recall, and F1-measure, respectively. Most subjects experience improvements, while only four subjects have decreased values. In particular, subjects 591 and 570 have significantly increased metrics, with the recall improving from 76\% to 94\% and from 79\% to 89\%, while the F1-measure from 86\% to 96\% and from 88\% to 94\%, respectively.
Similarly, improvements can be observed in class 1, while only two participants obtain decreased metrics. Notably, significant changes can be seen in subject 570, with the precision improving from 40\% to 44\%, the recall from 62\% to 72\%, and the F1-measure from 48\% to 55\%. In addition, the precision, recall, and F1-measure of subject 591 improves from 39\%, 53\%, and 45\% to 44\%, 78\%, and 56\% using individualization.
Thus, person-specific models result in the prediction of 10\% and 25\% more hypoglycemic cases up to 15 min before, respectively, for subjects 570 and 591. 
Finally, the minimum precision, recall, and F1-measure of the population-based models change from 25\%, 46\%, and 33\% to 20\%, 52\%, and 29\%, respectively for the subject-specific models, while the metrics are observed in different subjects. Similar outcomes can also be seen in class 2. Subjects 599, 570, 591, and 596 decreased in performance, whereas the remaining subjects had better metrics with the person-specific models. The most significant improvement can be observed in subject 540 with precision, recall, and F1-measure changing from 15\%, 57\%, and 23\% to 16\%, 64\%, and 26\%, respectively. The maximum values are the same and obtained for the same persons, while the lowest values remain similar but are noticed in different subjects. 
Class 3 does not reveal many changes. The best precision and F1-measure remain, while the recall improves from 44\% to 51\%. A significant decrease in recall performance is noticed in subjects 288 and 559.
Then, class 4 shows fewer variations, with three subjects improving and three subjects decreasing in performance with person-specific models. Similarly, the highest values are obtained for the same persons for precision and F1-measure but with decreased values compared to the population-based model, while the lowest values do not change. 
Class 5 induces similar outcomes. A significant change was noticed in subject 575, which had no correctly classified instances in the population model. The precision increases to 16\%, the recall to 41\%, and the F1-measure to 23\%. In addition, classes 6 and 7 present similar behavior, with most subjects showing slight increases in values while three subjects experience decreases. In class 8, performance improves in only three subjects, while the remaining test participants show declines. Overall, most subjects achieve better results using subject-specific models, particularly in the first classes and in recall.  
\\
Fig. \ref{fig:CMAll9} reveals the confusion matrices for the population-based and subject-specific models using LSTM and ResNet. The ResNet model has an increased recall, and the instances are better classified, especially in the first classes, using the personalized approach. 
\begin{table}[!ht]
\centering
\caption[Subject-Specific LSTM Results for Each Subject Using 9 Classes]{Subject-Specific LSTM Results for Each Subject Using 9 Classes \label{tab:LSTMIndi9}}
\setlength\extrarowheight{-2pt}
\resizebox{!}{7cm}
{\begin{tabular}
{|p{1.5cm}|p{1.5cm}|p{1.2cm}|p{1.2cm}|p{1.2cm}|p{1.2cm}|p{1.2cm}|p{1.2cm}|p{1.2cm}|p{1.2cm}|p{1.2cm}|}
\cline{1-11} \textbf{Subject} & \textbf{Metric} & 
\multicolumn{9}{ l |} {\textbf{Class}}\\ 
\cline{3-11} & & 0 & 1  & 2 & 3 & 4 & 5 & 6 & 7  & 8  \\ \hline
    
\multicolumn{1}{|l|}{\multirow{3}{*}{}}  & Precision & \textbf{1.00} &  0.35   &0.16  &0.11  & 0.12 & 0.13 & 0.40 & 0.21 & 0.39 \\ 

\cline{2-11} 
\multicolumn{1}{|l|}{\textbf{540}}   & Recall & \textbf{1.00} & 0.79 & 0.64     & 0.20 & 0.18 &0.11  & 0.26 & 0.19 & 0.24  \\ 

\cline{2-11} 
\multicolumn{1}{|l|}{} & F1-M &\textbf{1.00} & 0.49 & 0.26 & 0.14 & 0.14 & 0.12 & 0.31   & 0.20 & 0.30 \\  \hline\hline

\multicolumn{1}{|l|}{\multirow{6}{*}{}}  & Precision  &  0.97  & 0.20  & 0.12 & 0.09 & 0.00 & 0.00 & 0.24 & 0.23 & \textbf{0.60} \\ 

\cline{2-11}  
\multicolumn{1}{|l|}{\textbf{544}}   & Recall&  0.92&  0.54& 0.60     & 0.39 & 0.00 &0.00  & 0.48 & 0.19 & 0.42  \\ 

\cline{2-11} 
\multicolumn{1}{|l|}{} & F1-M& 0.94& 0.29 & 0.19 & 0.14& 0.00 & 0.00 &0.32    &0.21  & \textbf{0.50} \\  \hline\hline

\multicolumn{1}{|l|}{\multirow{6}{*}{}}  &Precision  &  \textbf{1.00}  & 0.59  & 0.15 &0.08  & 0.06 &  0.13&  0.17& 0.15 & 0.48 \\ 

\cline{2-11}  
\multicolumn{1}{|l|}{\textbf{552}}   & Recall& 0.93& 0.52 &\textbf{0.71}      &0.37  & 0.12 & 0.20 & 0.18 & 0.09 &0.25   \\ 

\cline{2-11} 
\multicolumn{1}{|l|}{} & F1-M& 0.96&  0.55&  0.25& 0.13 & 0.08 & 0.15 &0.17  &0.11    & 0.33 \\  \hline\hline

\multicolumn{1}{|l|}{\multirow{6}{*}{}}  &Precision  &  \textbf{1.00}&0.38     &0.16  &0.09  &0.08  & 0.12 & 0.33 & 0.17 &0.36  \\ 

\cline{2-11}  
\multicolumn{1}{|l|}{\textbf{559}}   &Recall& 0.99 &  0.65&   0.57   &0.16  &0.19  &0.20  & 0.10 & 0.29 &0.13   \\ 

\cline{2-11}  
\multicolumn{1}{|l|}{} & F1-M &\textbf{1.00} & 0.48 &0.25  &0.11  &0.12  & 0.15 & 0.16   & 0.21 & 0.19 \\  \hline\hline

\multicolumn{1}{|l|}{\multirow{6}{*}{}}  & Precision &   \textbf{1.00} & 0.43  & 0.16 & 0.13 & 0.16 & 0.12 & 0.02& 0.18 & 0.29 \\ 

\cline{2-11}  
\multicolumn{1}{|l|}{\textbf{563}}   & Recall& 0.99 & \textbf{0.82} &0.58      & 0.36 & \textbf{0.37} &  0.08& 0.01 & 0.18 & 0.15  \\ 

\cline{2-11} 
\multicolumn{1}{|l|}{} & F1-M & \textbf{1.00}& 0.56 &0.25  & 0.19 & 0.22 & 0.10 &0.02    &0.18  & 0.20 \\  \hline\hline

\multicolumn{1}{|l|}{\multirow{6}{*}{}}  & Precision  &\textbf{1.00}  & 0.44    & \textbf{0.27} & 0.19 &0.14  & 0.14 &  \textbf{0.42}&  0.28& 0.52 \\ 

\cline{2-11} 
\multicolumn{1}{|l|}{\textbf{567}}   & Recall& \textbf{1.00} & 0.75 & 0.67    &0.33  & 0.31 & 0.23 & 0.09 & \textbf{0.31} & \textbf{0.45}  \\ 

\cline{2-11} 
\multicolumn{1}{|l|}{} & F1-M & \textbf{1.00} & 0.55 & \textbf{0.39} &  0.24& 0.19 & 0.17 & 0.14   &\textbf{0.29}  &0.48  \\  \hline\hline

\multicolumn{1}{|l|}{\multirow{6}{*}{}}  & Precision  &  \textbf{1.00}  & 0.44  &0.13  &\textbf{0.20}  & \textbf{0.25} &\textbf{0.30}  &0.29  &\textbf{0.32}  &0.58  \\ 

\cline{2-11} 
\multicolumn{1}{|l|}{\textbf{570}}   & Recall& 0.89 &0.72  &0.18  & 0.33 & 0.29 & 0.15 & \textbf{0.68} & 0.16 &0.20   \\ 

\cline{2-11} 
\multicolumn{1}{|l|}{} & F1-M & 0.94& 0.55 &0.15  &\textbf{0.25}  & \textbf{0.27} & 0.20 & \textbf{0.41} & 0.21 & 0.30 \\  \hline\hline

\multicolumn{1}{|l|}{\multirow{6}{*}{}}  & Precision    & 0.98 & 0.33  & 0.14 & 0.15 & 0.14 & 0.16 & 0.22 & 0.18 & 0.10 \\ 

\cline{2-11} 
\multicolumn{1}{|l|}{\textbf{575}}   &Recall & 0.97  & 0.67 & 0.44  & \textbf{0.51} & 0.35 & \textbf{0.41} & 0.22 & 0.15 & 0.01   \\ 

\cline{2-11} 
\multicolumn{1}{|l|}{} & F1-M & 0.98& 0.44 & 0.21 & 0.23 & 0.20  & \textbf{0.23}   &0.22  & 0.16 & 0.02 \\  \hline\hline

\multicolumn{1}{|l|}{\multirow{6}{*}{}}  & Precision  &  \textbf{1.00}& \textbf{1.00} & \textbf{0.27}  & 0.10 & 0.15 & 0.06 & 0.08 & 0.02 & 0.45 \\ 

\cline{2-11} 
\multicolumn{1}{|l|}{\textbf{584}}   & Recall &  \textbf{1.00}&  0.75 & 0.67  & 0.28 & 0.25 & 0.07 & 0.17 & 0.01 & 0.27  \\ 

\cline{2-11} 
\multicolumn{1}{|l|}{} & F1-M & \textbf{1.00}& \textbf{0.86} &\textbf{0.39}  & 0.14 & 0.19 & 0.06 & 0.11  & 0.01 &0.34  \\  \hline\hline

\multicolumn{1}{|l|}{\multirow{6}{*}{}}  & Precision  &   0.98 & 0.33  &0.08  &0.07  & 0.14 & 0.16 & 0.38 & 0.09 & 0.42 \\ 

\cline{2-11} 
\multicolumn{1}{|l|}{\textbf{588}}   & Recall & \textbf{1.00} & 0.56 & 0.29 & 0.08 & 0.17 & 0.25 & 0.42 & 0.01 & 0.30  \\ 

\cline{2-11} 
\multicolumn{1}{|l|}{} & F1-M & 0.99&  0.41&  0.13&  0.08& 0.15 & 0.20 & 0.40   &0.01  &0.35  \\  \hline\hline

\multicolumn{1}{|l|}{\multirow{6}{*}{}}  & Precision &  0.99  & 0.44  &0.10  & 0.06 & 0.02 & 0.10 & 0.32 & 0.12 & 0.34 \\ 

\cline{2-11}  
\multicolumn{1}{|l|}{\textbf{591}}   & Recall & 0.94 & 0.78   & 0.36  &0.20  &0.04  & 0.10 & 0.10 & 0.12&0.33   \\ 

\cline{2-11} 
\multicolumn{1}{|l|}{} & F1-M& 0.96&  0.56& 0.15 & 0.10 &0.03  &0.10    & 0.16 &0.12 & 0.33 \\  \hline\hline

\multicolumn{1}{|l|}{\multirow{6}{*}{}}  & Precision  &  \textbf{1.00} & 0.52  & 0.18 & 0.09 & 0.14 & 0.13 & 0.06 & 0.17 & 0.45 \\ 

\cline{2-11} 
\multicolumn{1}{|l|}{\textbf{596}}   & Recall & \textbf{1.00} & \textbf{0.82} &  0.40    & 0.45 & 0.19 & 0.23 & 0.01 & 0.21 & 0.28  \\ 

\cline{2-11} 
\multicolumn{1}{|l|}{} & F1-M & \textbf{1.00}& 0.64 & 0.25 & 0.15 & 0.16 & 0.17 & 0.01 &0.19 & 0.34 \\  \hline

\end{tabular}}
\end{table}

\begin{table}[!ht]
\centering
\caption[Population-Based LSTM Results with Less Test Data for Each Subject Using 9 Classes]{Population-Based LSTM Results with Less Test Data for Each Subject Using 9 Classes \label{tab:LSTMPop9less}}
\setlength\extrarowheight{-2pt}
\resizebox{!}{7cm}
{\begin{tabular}
{|p{1.5cm}|p{1.5cm}|p{1.2cm}|p{1.2cm}|p{1.2cm}|p{1.2cm}|p{1.2cm}|p{1.2cm}|p{1.2cm}|p{1.2cm}|p{1.2cm}|}
\cline{1-11} \textbf{Subject} & \textbf{Metric} & 
\multicolumn{9}{ l |} {\textbf{Class}}\\ 
\cline{3-11} & & 0 & 1  & 2 & 3 & 4 & 5 & 6 & 7  & 8  \\ \hline
    
\multicolumn{1}{|l|}{\multirow{3}{*}{}}  & Precision & \textbf{1.00} &  0.37   &0.15  &0.11  & 0.14 & 0.13 & 0.35 & 0.25 & 0.43 \\ 

\cline{2-11} 
\multicolumn{1}{|l|}{\textbf{540}}   & Recall & 0.99 & 0.76 & 0.57     & 0.19 & 0.25 &0.12  & 0.21 & 0.28 & 0.23  \\ 

\cline{2-11} 
\multicolumn{1}{|l|}{} & F1-M &\textbf{1.00} & 0.50 & 0.23 & 0.14 & 0.18 & 0.12 & 0.27   & 0.26 & 0.30 \\  \hline\hline

\multicolumn{1}{|l|}{\multirow{6}{*}{}}  & Precision  &  \textbf{1.00}& 0.26  & 0.11 & 0.10 & 0.00 & 0.00 & 0.21 & 0.50 & \textbf{0.65} \\ 

\cline{2-11}  
\multicolumn{1}{|l|}{\textbf{544}}   & Recall&  0.81&  0.46& 0.60     & 0.33 & 0.00 &0.00  & 0.45 & 0.18 & \textbf{0.59}  \\ 

\cline{2-11} 
\multicolumn{1}{|l|}{} & F1-M& 0.90& 0.33 & 0.18 & 0.16& 0.00 & 0.00 &0.29    &0.26  & \textbf{0.62} \\  \hline\hline

\multicolumn{1}{|l|}{\multirow{6}{*}{}}  &Precision  &  \textbf{1.00}  & 0.52  & 0.14 &0.08  & 0.05 &  0.15&  0.14& 0.09 & 0.59 \\ 

\cline{2-11}  
\multicolumn{1}{|l|}{\textbf{552}}   & Recall& 0.95& 0.52 &\textbf{0.71}      &0.40  & 0.08 & \textbf{0.32} & 0.13 & 0.05 &0.27   \\ 

\cline{2-11} 
\multicolumn{1}{|l|}{} & F1-M& 0.97&  0.52&  0.24& 0.14 & 0.06 & \textbf{0.20} &0.13  &0.06    & 0.36 \\  \hline\hline

\multicolumn{1}{|l|}{\multirow{6}{*}{}}  &Precision  & \textbf{1.00}&0.39    &0.19  &0.13  &0.10  & 0.19 & 0.36 & 0.22 &0.46  \\ 

\cline{2-11}  
\multicolumn{1}{|l|}{\textbf{559}}   &Recall& 0.97 &  0.68&   0.68   &0.27  &0.26  &0.20  & 0.21 & 0.25 &0.27   \\ 

\cline{2-11}  
\multicolumn{1}{|l|}{} & F1-M &0.98 & 0.49 &0.29  &0.17  &0.14  & 0.19 & 0.26   & 0.24 & 0.34 \\  \hline\hline

\multicolumn{1}{|l|}{\multirow{6}{*}{}}  & Precision &   \textbf{1.00} & 0.42  & 0.16 & 0.13 & 0.16 & 0.12 & 0.02& 0.18 & 0.28 \\ 

\cline{2-11}  
\multicolumn{1}{|l|}{\textbf{563}}   & Recall& 0.98 & 0.78 &0.58      & 0.36 & 0.37 &  0.07& 0.02 & 0.18 & 0.15  \\ 

\cline{2-11} 
\multicolumn{1}{|l|}{} & F1-M & 0.99& 0.54 &0.25  & 0.19 & 0.22 & 0.09 &0.02    &0.18  & 0.20 \\  \hline\hline

\multicolumn{1}{|l|}{\multirow{6}{*}{}}  & Precision  &\textbf{1.00}  & 0.43    & 0.25 & \textbf{0.20} &0.13  & 0.12 &  \textbf{0.45}&  0.28& 0.49\\ 

\cline{2-11} 
\multicolumn{1}{|l|}{\textbf{567}}   & Recall& \textbf{1.00} & 0.71 & 0.63    &0.33  & 0.31 & 0.19 & 0.08 & \textbf{0.30} & 0.44  \\ 

\cline{2-11} 
\multicolumn{1}{|l|}{} & F1-M & \textbf{1.00} & 0.53 & 0.36 & \textbf{0.25}& 0.19 & 0.15 & 0.13   &\textbf{0.29}  &0.46  \\  \hline\hline

\multicolumn{1}{|l|}{\multirow{6}{*}{}}  & Precision  &  \textbf{1.00}  & 0.40  &0.18  &0.14  & \textbf{0.30} &\textbf{0.30}  &0.27  &0.31  &0.55  \\ 

\cline{2-11} 
\multicolumn{1}{|l|}{\textbf{570}}   & Recall& 0.79 &0.62  &0.24  & 0.21 & 0.29 & 0.15 & \textbf{0.67} & 0.17 &0.18   \\ 

\cline{2-11} 
\multicolumn{1}{|l|}{} & F1-M & 0.88& 0.48 &0.20  &0.17  & \textbf{0.30} & \textbf{0.20} & \textbf{0.38} & 0.22 & 0.28 \\  \hline\hline

\multicolumn{1}{|l|}{\multirow{6}{*}{}}  & Precision    & 0.93 & 0.25  & 0.13 & 0.13 & 0.12 & 0.00 & 0.18 & 0.31 & 0.39 \\ 

\cline{2-11} 
\multicolumn{1}{|l|}{\textbf{575}}   &Recall & \textbf{1.00}  & 0.71 & 0.46  & 0.29 &\textbf{0.42} & 0.00 & 0.18 & 0.04 & 0.43   \\ 

\cline{2-11} 
\multicolumn{1}{|l|}{} & F1-M & 0.97& 0.37 & 0.20 & 0.18 & 0.19  &0.00  &0.18  & 0.08 & 0.41 \\  \hline\hline

\multicolumn{1}{|l|}{\multirow{6}{*}{}}  & Precision  &  \textbf{1.00}& \textbf{1.00} & \textbf{0.27}  & 0.10 & 0.16 & 0.06 & 0.08 & 0.02 & 0.45 \\ 

\cline{2-11} 
\multicolumn{1}{|l|}{\textbf{584}}   & Recall &  \textbf{1.00}&  0.75 & 0.67  & 0.28 & 0.25 & 0.07 & 0.17 & 0.02 & 0.28  \\ 

\cline{2-11} 
\multicolumn{1}{|l|}{} & F1-M & \textbf{1.00}& \textbf{0.86} &\textbf{0.39}  & 0.14 & 0.19 & 0.06 & 0.11  & 0.02 &0.34  \\  \hline\hline

\multicolumn{1}{|l|}{\multirow{6}{*}{}}  & Precision  &   0.95 & 0.41  &0.07 &0.13  & 0.11 & 0.17 & 0.29 & \textbf{0.80} & 0.19 \\ 

\cline{2-11} 
\multicolumn{1}{|l|}{\textbf{588}}   & Recall & \textbf{1.00} & 0.60 & 0.17 & 0.21 & 0.34 & 0.14 & 0.38 & 0.05 & 0.06  \\ 

\cline{2-11} 
\multicolumn{1}{|l|}{} & F1-M & 0.97&  0.48&  0.10&  0.16& 0.16 & 0.16 & 0.33   &0.10  &0.09  \\  \hline\hline

\multicolumn{1}{|l|}{\multirow{6}{*}{}}  & Precision &  \textbf{1.00}  & 0.39  &0.13  & 0.09 & 0.11 & 0.08 & 0.36 & 0.17 & 0.43 \\ 

\cline{2-11}  
\multicolumn{1}{|l|}{\textbf{591}}   & Recall & 0.76 & 0.53   & 0.38  &0.35  &0.09  & 0.12 & 0.26 & 0.11&0.39   \\ 

\cline{2-11} 
\multicolumn{1}{|l|}{} & F1-M& 0.86&  0.45& 0.20 & 0.14 &0.10  &0.10    & 0.30 &0.13 & 0.41 \\  \hline\hline

\multicolumn{1}{|l|}{\multirow{6}{*}{}}  & Precision  &  \textbf{1.00} & 0.55  & 0.21 & 0.10 & 0.16 & 0.15 & 0.00 & 0.18 & 0.41 \\ 

\cline{2-11} 
\multicolumn{1}{|l|}{\textbf{596}}   & Recall & 0.99& \textbf{0.79} &  0.40    & \textbf{0.44} & 0.18 & 0.16 & 0.00 & 0.26 & 0.36  \\ 

\cline{2-11} 
\multicolumn{1}{|l|}{} & F1-M & 0.99& 0.65 & 0.27 & 0.16 & 0.17 & 0.16 & 0.00 &0.21 & 0.38 \\  \hline
\end{tabular}}
\end{table}

\begin{table}[h]
\centering
\caption[Comparison of Population-Based and Subject-Specific Approaches for Each Model Using 9 Classes]{Comparison of Population-Based and Subject-Specific Approaches for Each Model Using 9 Classes
\label{tab:Comp9classlesstest}}
\setlength\extrarowheight{-2pt}
\resizebox{!}{4.5cm}
{\begin{tabular}{p{0.5cm}|p{1.5cm}|p{1.2cm}|p{1.2cm}|p{1.2cm}|p{1.2cm}|p{1.2cm}|p{1.2cm}|p{1.2cm}|p{1.2cm}|p{1.2cm}|p{1.2cm}|}
\cline{2-12}
    & \textbf{Metrics} & \textbf{Avg} &
    \multicolumn{9}{ l |} {\textbf{Classes}}\\ 
    \cline{4-12} & & & 0 & 1  & 2 & 3 & 4 & 5 & 6 & 7  & 8  \\ \hline

\cline{2-12} 
\multicolumn{1}{|l|}{\multirow{4}{*}{}} & Precision & 0.28 &0.94 &0.33  &0.15&0.11&0.10& 0.11&0.19&0.18 & 0.37    \\  

\cline{2-12}  
\multicolumn{1}{|l|}{\textbf{PB ResNet}} & Recall  &0.32  &0.90 &0.61  &0.34&0.17&0.17 &0.08& 0.20& 0.31  &0.17   \\ 

\cline{2-12}  
\multicolumn{1}{|l|}{}  & F1-M  & 0.28 &0.92 & 0.43 &0.21&0.14& 0.10&0.09 &0.19   &0.23 & 0.24 \\   
 \hline\hline

\multicolumn{1}{|l|}{\multirow{4}{*}{}} &  Precision & 0.30 &0.92 &0.28  &0.14&0.13&0.13 &0.17&0.29 &0.21   &0.45  \\    

\cline{2-12}  
\multicolumn{1}{|l|}{\textbf{SS ResNet}} & Recall &0.38  &0.91 & 0.65 &0.43&0.31& 0.26&0.21& 0.26& 0.19  &0.21  \\

\cline{2-12}  
\multicolumn{1}{|l|}{}  & F1-M  & 0.31 &0.92 &0.39  &0.21&0.18&0.18 &0.19&0.27 &0.20   &0.29 \\
\hline\hline

\multicolumn{1}{|l|}{\multirow{4}{*}{}} & Precision & 0.31 &0.99 & 0.39 &0.16&0.11&0.13 &0.14& 0.24& 0.21  &0.45  \\ 

\cline{2-12}  
\multicolumn{1}{|l|}{\textbf{PB LSTM}}    & Recall  & 0.39 &0.95 & 0.68 &0.49&0.29& 0.25&0.13& 0.20&0.16   &0.33 \\  

\cline{2-12}  
\multicolumn{1}{|l|}{} & F1-M  &0.33  &0.97 & 0.49 &0.24&0.16& 0.15&0.13& 0.22& 0.18  & 0.38 \\    
\hline\hline

\multicolumn{1}{|l|}{\multirow{4}{*}{}} & Precision & 0.31 &1.00 & 0.40 &0.15&0.11&0.12 &0.14& 0.25& 0.18  &0.48  \\ 

\cline{2-12}  
\multicolumn{1}{|l|}{\textbf{SS LSTM}}   & Recall & 0.39 &0.98 &0.72  &0.50&0.29&0.21 &0.18&0.19 &0.17   &0.24  \\  

\cline{2-12}  
\multicolumn{1}{|l|}{}  & F1-M  &  0.32& 0.99& 0.52 & 0.23&0.16&0.15&0.16 &0.21& 0.17&0.31    \\  
\hline\hline

\multicolumn{1}{|l|}{\multirow{4}{*}{}} & Precision & 0.28 &0.95 & 0.26 &0.15&0.14&0.11 &0.16&0.22 &0.23   &0.35  \\

\cline{2-12}  
\multicolumn{1}{|l|}{\textbf{PB Hybrid}}   & Recall & 0.33 &0.80 & 0.47 &0.38&0.30&0.24&0.20 &0.07   & 0.34 & 0.21  \\  

\cline{2-12}  
\multicolumn{1}{|l|}{} & F1-M  &  0.29&0.87 &0.33  &0.21&0.19&0.15 &0.18&0.11 &0.28   &0.26 \\     
\hline\hline

\multicolumn{1}{|l|}{\multirow{4}{*}{}} & Precision & 0.28 &0.92 &0.26  &0.12&0.11&0.11 &0.19&0.22 &0.23   &0.36  \\ 

\cline{2-12}  
\multicolumn{1}{|l|}{\textbf{SS Hybrid}}    & Recall  &0.36&0.89  &0.59 & 0.37 &0.31&0.23& 0.23&0.07&0.33 &0.19     \\  

\cline{2-12}  
\multicolumn{1}{|l|}{} & F1-M  & 0.29& 0.91 &0.36 & 0.18 &0.16&0.15&0.21 &0.11& 0.27&0.25     \\  \hline
\end{tabular}}
\end{table}
\vspace{-2em}

\subsection{Population-Based Models Using 6 Classes}
\label{sub:Pop6class}

The comparison in the previous subsections showed that the proposed models cannot differentiate between the later classes. In contrast, the earlier classes performed better, indicating that approximately 60-70\% of all events can be predicted, but with a low precision of 55\%, as can be seen in Table \ref{tab:Comp6class}. To address this, the same models were tested using only six classes, focusing on predictions up to 4h before hypoglycemia. Based on the macro average metrics for each subject and model, the LSTM model still demonstrates superior performance. On average, approximately 70\% of all events are classified, however the F1-score remains below 65\%. The lowest best metrics include a precision of 51\%, a recall of 49\%, and an F1-score of 49\% using an LSTM model. Therefore, at worst, around 50\% of instances are classified. On average, considering the highest values for each subject, a precision of 55\%, a recall of 58\%, and an F1-score of 55\% are achieved. 
\\
Looking at Table \ref{tab:LSTMPop6}, which reveals the performance of single classes obtained with the LSTM model, high variations between the subjects can be seen. Similarly, class 0 is distinctive, with an average classification performance of 99\% and a minimum of at least 94\%, which is an increase of 14\%. For class 1, more variations are evident. Subject 567 reports the best result with a precision of 66\% and an F1-measure of 70\%, while subject 552 has the best recall with 78\% but a precision of only 61\%. Compared to the results obtained with nine classes, it can be seen that the precision improves. The worst metrics are a precision of 46\%, a recall of 52\%, and an F1-measure of 49\%.
Moving to class 2, subject 567 still has the best results with increased performance, while the precision is still low. Increases in metrics are seen in almost all subjects, while some only show improvement in precision and F1-measure.
The population performance of class 2 across all subjects is 27\%, 53\%, and 36\% for precision, recall, and F1-measure, respectively, which varies significantly from the performance of class 1. The worst performance can be seen with 17\%, 37\%, and 23\% in subject 584, who had even decreased performance with nine classes. For classes 3 and 4, the results decline even more with the best recall, precision, and F1-measure of 42-44\%, 43-48\%, and 41-42\%, respectively. The average F1-measure improves from 18\% to 31\% for class 3 and from 18\% to 34\% for class 4. Lastly, a significant improvement is observed in class 5, where the population F1-measure across all subjects increases from 15\% with nine classes to 69\% with six classes. The best results are achieved with precision, recall, and F1-measure values of 80\%, 79\%, and 80\%, respectively. Conversely, the worst performance is recorded with a precision of 63\%, recall of 48\%, and an F1-measure of 55\%. Overall, using less classes stabilizes performance, nevertheless the classification ability remains insufficient. Subject 567 achieves the highest average metric values across all classes, whereas the worst performance is observed in subjects 584 and 544, with lower precision and recall compared to other patients. However, even with the reduced classes, the model cannot differentiate between classes 3 and 4. In contrast, class 5 shows a significant improvement. 
\\
The confusion matrices of subject 567 for the LSTM model are presented in \ref{fig:CMPop9}. Likewise, it can be seen that most misclassifications are within the neighbor classes, especially for classes 0, 1, and 2. 

\begin{table}[!ht]
\centering
\caption[LSTM Results for Each Subject Using 6 Classes]{LSTM Results for Each Subject Using 6 Classes \label{tab:LSTMPop6}}
\setlength\extrarowheight{-2pt}
\resizebox{!}{7cm}
{\begin{tabular}
{|p{1.5cm}|p{2cm}|p{1.8cm}|p{1.8cm}|p{1.8cm}|p{1.8cm}|p{1.8cm}|p{1.8cm}|}
\cline{1-8}\textbf{Subject} & \textbf{Metric} & 
\multicolumn{6}{ l |} {\textbf{Class}}\\ 
\cline{3-8} & & 0 & 1  & 2 & 3 & 4 & 5   \\ \hline
    
\multicolumn{1}{|l|}{\multirow{3}{*}{}}  & Precision & \textbf{1.00} & 0.60    &0.28  & 0.26 &  0.38& 0.69 \\ 

\cline{2-8}
\multicolumn{1}{|l|}{\textbf{540}}   & Recall &0.99 & 0.72& 0.52&0.32 &0.31 & 0.59  \\ 

\cline{2-8}
\multicolumn{1}{|l|}{} & F1-M & 0.99& 0.66& 0.37& 0.29& 0.34& 0.64\\  \hline\hline

\multicolumn{1}{|l|}{\multirow{6}{*}{}}  & Precision  &0.96 &0.46 &0.25 &0.21 &0.43 & 0.79 \\ 

\cline{2-8}
\multicolumn{1}{|l|}{\textbf{544}}   & Recall&0.94 &0.52 & 0.37& 0.23& 0.35& 0.76   \\ 

\cline{2-8}
\multicolumn{1}{|l|}{} & F1-M&0.95 &0.49 &0.30 &0.22 &0.38 &0.78  \\  \hline\hline

\multicolumn{1}{|l|}{\multirow{6}{*}{}}  &Precision & \textbf{1.00}& 0.61& 0.26& 0.24& 0.36& 0.74 \\ 

\cline{2-8}
\multicolumn{1}{|l|}{\textbf{552}}   & Recall&0.99&\textbf{0.78} &0.62 &0.28 & 0.28&0.62\\ 

\cline{2-8}
\multicolumn{1}{|l|}{} & F1-M& 0.99&0.69 &0.36 &0.26 & 0.32&0.67  \\  \hline\hline

\multicolumn{1}{|l|}{\multirow{6}{*}{}}  &Precision  &\textbf{1.00} &0.62 & 0.27& 0.26& 0.37& 0.73 \\ 

\cline{2-8} 
\multicolumn{1}{|l|}{\textbf{559}}   &Recall& \textbf{1.00}&0.70 &0.66 &0.37 &0.23 &0.64  \\ 

\cline{2-8} 
\multicolumn{1}{|l|}{} & F1-M &\textbf{1.00} &0.66 &0.38 &0.31 &0.28 & 0.68 \\  \hline\hline

\multicolumn{1}{|l|}{\multirow{6}{*}{}}  & Precision& \textbf{1.00}& 0.63& 0.35& 0.36& \textbf{0.48}& 0.75  \\ 

\cline{2-8} 
\multicolumn{1}{|l|}{\textbf{563}}   & Recall&0.99 &0.74 & 0.55& 0.40& 0.34& 0.76   \\ 

\cline{2-8}
\multicolumn{1}{|l|}{} & F1-M &0.99 &0.68 &0.43 &0.38 & 0.40& 0.75 \\  \hline\hline

\multicolumn{1}{|l|}{\multirow{6}{*}{}}  & Precision &\textbf{1.00} &\textbf{0.66} &\textbf{0.41} &\textbf{0.43} &0.45 &\textbf{0.80}   \\ 

\cline{2-8}
\multicolumn{1}{|l|}{\textbf{567}}   & Recall& \textbf{1.00}& 0.75& \textbf{0.67}& \textbf{0.42}& 0.37&\textbf{0.79}   \\ 

\cline{2-8}
\multicolumn{1}{|l|}{} & F1-M & \textbf{1.00}& \textbf{0.70}&\textbf{0.51} &\textbf{0.42} &\textbf{0.41} &\textbf{0.80}   \\  \hline\hline

\multicolumn{1}{|l|}{\multirow{6}{*}{}}  & Precision  &0.99 &0.49 &0.19 &0.34 &0.39 &0.75  \\ 

\cline{2-8}
\multicolumn{1}{|l|}{\textbf{570}}   & Recall&0.98 & 0.64& 0.38& 0.31& 0.27&0.72  \\ 

\cline{2-8} 
\multicolumn{1}{|l|}{} & F1-M &0.99 &0.59 &0.25 &0.32 &0.32 &0.73   \\  \hline\hline

\multicolumn{1}{|l|}{\multirow{6}{*}{}}  & Precision &0.99 &0.64 &0.27 &0.33 & 0.45& 0.77 \\ 

\cline{2-8}
\multicolumn{1}{|l|}{\textbf{575}}   &Recall&0.99 &0.73 & 0.52& 0.34& 0.36& 0.71  \\ 

\cline{2-8} 
\multicolumn{1}{|l|}{} & F1-M &0.99 & 0.68&0.35 &0.33 &0.40 &0.74  \\  \hline\hline

\multicolumn{1}{|l|}{\multirow{6}{*}{}}  & Precision &\textbf{1.00} &0.56 &0.17 & 0.21&0.31 &0.63  \\ 

\cline{2-8}
\multicolumn{1}{|l|}{\textbf{584}}   & Recall &\textbf{1.00}& 0.65& 0.37&0.27 &0.28 & 0.48  \\ 

\cline{2-8}
\multicolumn{1}{|l|}{} & F1-M & \textbf{1.00}&0.60 & 0.23& 0.24& 0.29& 0.55  \\  \hline\hline

\multicolumn{1}{|l|}{\multirow{6}{*}{}}  & Precision &\textbf{1.00} &0.59 &0.21 &0.27 &0.37 & 0.75  \\ 

\cline{2-8}
\multicolumn{1}{|l|}{\textbf{588}}   & Recall & 0.99& 0.66& 0.40& 0.31& \textbf{0.44}&0.53  \\ 

\cline{2-8}
\multicolumn{1}{|l|}{} & F1-M & \textbf{1.00} & 0.63& 0.27& 0.28& 0.40& 0.62 \\  \hline\hline

\multicolumn{1}{|l|}{\multirow{6}{*}{}}  & Precision &0.98 &0.61 &0.27 &0.28 &0.30 &0.66  \\ 

\cline{2-8}
\multicolumn{1}{|l|}{\textbf{591}}   & Recall &\textbf{1.00} &0.76 & 0.48& 0.35& 0.25& 0.55  \\ 

\cline{2-8}
\multicolumn{1}{|l|}{} & F1-M& 0.99& 0.68& 0.34& 0.31& 0.27&0.60 \\  \hline\hline

\multicolumn{1}{|l|}{\multirow{6}{*}{}}  & Precision &\textbf{1.00} &0.57 &0.30 &0.30 &0.37 &0.77  \\ 

\cline{2-8}
\multicolumn{1}{|l|}{\textbf{596}}   & Recall &0.97 &0.63 &0.55 &0.37 &0.32 & 0.67  \\ 

\cline{2-8}
\multicolumn{1}{|l|}{} & F1-M & 0.98& 0.60& 0.39& 0.33&0.34 &0.72  \\  \hline

\end{tabular}}
\end{table}

\begin{table}[!ht]
\centering
\caption[Performance of Each Class for Each Model Using 6 Classes]{Performance of Each Class for Each Model Using 6 Classes \label{tab:Comp6class}}
\setlength\extrarowheight{-2pt}
 \resizebox{!}{3cm}
{\begin{tabular}{p{0.5cm}|p{1.6cm}|p{0.9cm}|p{0.9cm}|p{0.9cm}|p{0.9cm}|p{0.9cm}|p{0.9cm}|p{0.9cm}|}
\cline{2-9}
    & \textbf{Metric} & \textbf{Avg} &
    \multicolumn{6}{ l |} {\textbf{Class}}\\ 
    \cline{4-9} & & & 0& 1  & 2 & 3 & 4 & 5 \\ \hline

\multicolumn{1}{|l|}{\multirow{4}{*}{}} & Precision & 0.51  & 0.95  & 0.47 &0.27&0.29 &0.37  &0.68  \\  

\cline{2-9} 
\multicolumn{1}{|l|}{\textbf{ResNet}} & Recall  &  0.53 &  0.96 & 0.60 & 0.38&0.31& 0.23 & 0.73  \\ 

\cline{2-9}
\multicolumn{1}{|l|}{}  & F1-M  & 0.51 &  0.95& 0.53  & 0.32 & 0.30 & 0.28 & 0.71  \\   
 \hline\hline

\multicolumn{1}{|l|}{\multirow{4}{*}{}} & Precision  & 0.55& 0.99 & 0.60  & 0.27 & 0.29 & 0.39 & 0.74\\ 

\cline{2-9} 
\multicolumn{1}{|l|}{\textbf{LSTM}}   & Recall  & 0.59  &0.99& 0.71  & 0.53 &0.34 & 0.31 & 0.65  \\  

\cline{2-9} 
\multicolumn{1}{|l|} {} & F1-M  &  0.56 & 0.99  & 0.65 & 0.36& 0.31 &0.34& 0.69 \\  \hline\hline

\multicolumn{1}{|l|}{\multirow{4}{*}{}} & Precision  &  0.51 & 0.96  & 0.46 & 0.26& 0.29 & 0.37 &0.70  \\

\cline{2-9}
\multicolumn{1}{|l|}{\textbf{Hybrid}}   & Recall  &  0.55 & 0.95  & 0.69  & 0.41 & 0.31 & 0.26&0.68  \\  

\cline{2-9} 
\multicolumn{1}{|l|}{} & F1-M  &  0.52 &  0.96 & 0.55& 0.32& 0.30& 0.31 & 0.69 \\     \hline
\end{tabular}}
\end{table}

\subsection{Subject-Specific and Population-Based Models Using 6 Classes}
\label{sub:6compare}

\begin{figure}[!ht]
	\centering
    \captionsetup{justification=centering}
    \subfloat[][]{\includegraphics[width=0.33\textwidth]{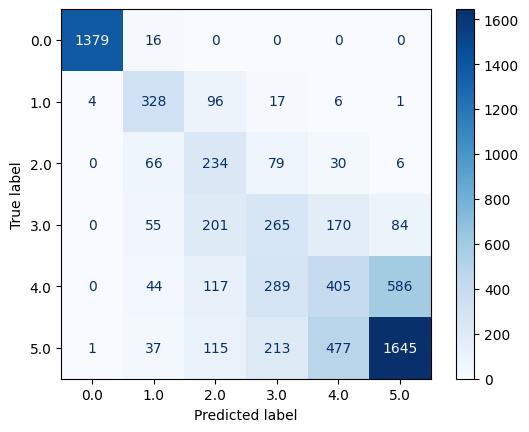}}
    \hspace{0.3cm}
    \subfloat[][]{\includegraphics[width=0.4\textwidth]{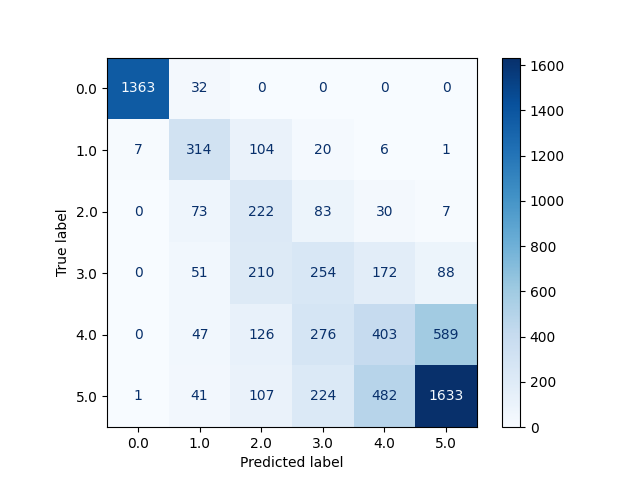}}
    \hspace{0.3cm}
    \subfloat[][]{\includegraphics[width=0.33\textwidth]{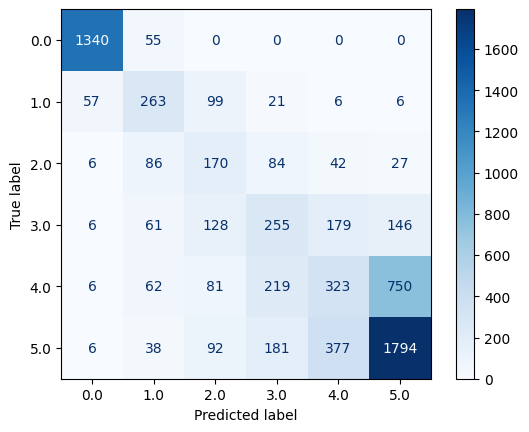}}
    \hspace{0.3cm}
    \subfloat[][]{\includegraphics[width=0.4\textwidth]{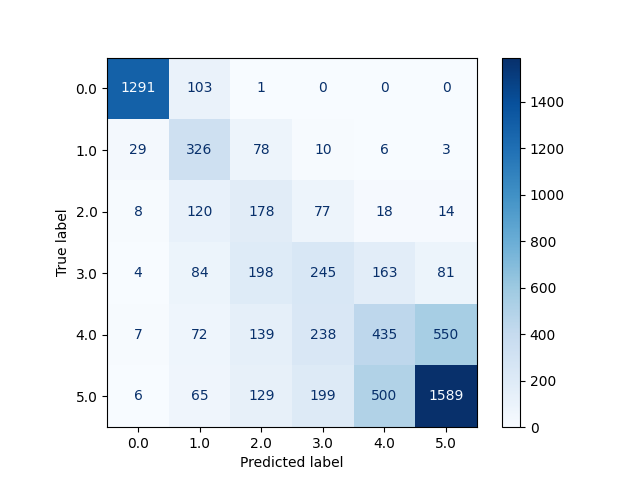}}
    \hspace{0.3cm}
    \subfloat[][]{\includegraphics[width=0.33\textwidth]{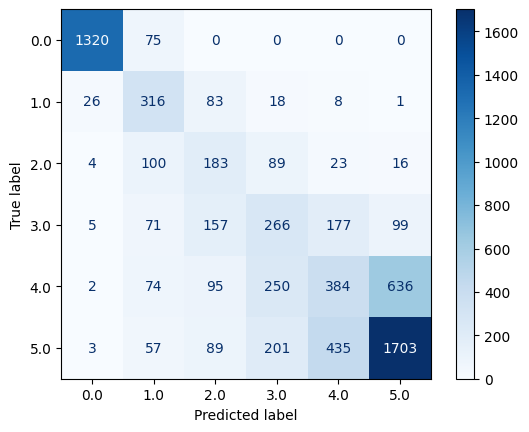}}
    \hspace{0.3cm}
    \subfloat[][]{\includegraphics[width=0.4\textwidth]{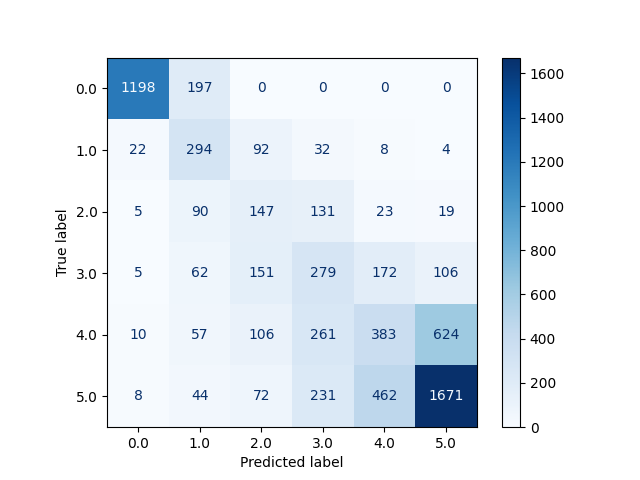}}
    \hspace{0.3cm}
    \caption[Population-Based and Subject-Specific Confusion-Matrices Across 6 Classes]{Population-Based and Subject-Specific Confusion-Matrices Across 6 Classes \label{fig:CMAll6}}
	 (a) Population-Based LSTM Model (b) Subject-Specific LSTM Model (c) Population-Based ResNet Model (d) Subject-Specific ResNet Model (e) Population-Based Hybrid Model (f) Subject-Specific Hybrid Model
\end{figure}

From Table \ref{tab:Comp6classless}, it is evident that both LSTM-based approaches yield similar performance across all subjects. However, the population-based model performs better in classes 1, 2, and 4, whereas the individualized model shows an advantage only in class 4. Comparing the macro averages across all models, both LSTM approaches achieve the best results, followed by the hybrid population-based model. The individualized models do not show significant differences. Notably, the individualized hybrid model experiences a decline in all metrics for the first classes and only outperforms in class 4, likewise the ResNet model. On average, training with six classes leads to more correctly classified instances using a population-based model. The LSTM model successfully classifies 60\% of all hypoglycemic events at the correct time, while 73\% of events can be predicted up to 15 minutes before occurrence with a 40\% false alarm rate.
\\
Thus, again, the performance of the population-based and subject-specific LSTM models can be seen in Tables \ref{tab:LSTMPop6less} and \ref{tab:LSTMIndi6}, respectively. This time, a significant improvement cannot be observed. For class 0, the performance remains the same for most patients, while three subjects have better values and two subjects experience a decline. The maximum values are obtained in five patients with 100\% for all metrics in the subject-specific and population-based model. Turning now to class 1, only the performance of six subjects improves, while the metrics of the other subjects either do not change or decrease. Notably, the recall of subject 567 improved from 77\% to 81\%, but the precision and F1-measure decreased. The maximum values are the same for both approaches, with 100\%, 83\%, and 91\% for precision, recall, and F1-measure, respectively, for the same subject. Likewise, in class 2, only three subjects improve in performance from an F1-measure of 44\% to 49\%, 39\% to 42\%, and 19\% to 25\%, respectively. The maximum values also do not change with 74\% and 49\% for recall and F1-measure, respectively. The best precision is increased by 1\% through individualization. Contrariwise, the metrics of the three subjects decrease. Similar behavior can be observed in class 3 as well. Two subjects increase in performance, two subjects decrease, while the remaining subjects do not experience significant changes. While the best precision is the same, recall and F1-measure decrease in the subject-specific model. In class 4, the maximum values increase using the subject-specific model from 58\%, 34\%, 42\% to 59\%, 41\%, and 44\% for precision, recall, and F1-measure, respectively. 
In class 5, almost all subjects behave similarly, with only a few variations. 
\\
Lastly, Fig. \ref{fig:CMAll6} shows the confusion matrices for each population-based and subject-specific model reflecting the observed behaviors.
\\
Conclusively, the proposed models cannot outperform the results of presented state-of-the-art studies. The same difficulties are faced, and the precision is very low. Therefore, it is noticed that a classification system with multiple classes causes more misclassifications and worse recall. Nevertheless, as pointed out, the confusion matrices reveal that most of the miss-classifications are with the nearest neighbors when using six classes and up to 30 min before the event can be classified with a good recall.

\begin{table}
\centering
\caption[Subject-Specific LSTM Results for Each Subject Using 6 Classes]{Subject-Specific LSTM Results for Each Subject Using 6 Classes \label{tab:LSTMIndi6}}
\setlength\extrarowheight{-2pt}
\resizebox{!}{7cm}
{\begin{tabular}
{|p{1.5cm}|p{2cm}|p{1.8cm}|p{1.8cm}|p{1.8cm}|p{1.8cm}|p{1.8cm}|p{1.8cm}|}
\cline{1-8}\textbf{Subject} & \textbf{Metric} & 
\multicolumn{6}{ l |} {\textbf{Class}}\\ 
\cline{3-8} & & 0 & 1  & 2 & 3 & 4 & 5   \\ \hline
    
\multicolumn{1}{|l|}{\multirow{3}{*}{}}  & Precision & \textbf{1.00} & 0.55   &0.30  & 0.30 &  0.40& 0.66 \\ 

\cline{2-8}
\multicolumn{1}{|l|}{\textbf{540}}   & Recall & \textbf{1.00}& 0.79& 0.54& 0.32& 0.35&0.55   \\ 

\cline{2-8}
\multicolumn{1}{|l|}{} & F1-M &\textbf{1.00} &0.65&0.39 & 0.31& 0.37&0.60  \\  \hline\hline

\multicolumn{1}{|l|}{\multirow{6}{*}{}}  & Precision  & 0.98& 0.48&\textbf{0.38} &0.25 & 0.39& 0.68 \\ 

\cline{2-8}
\multicolumn{1}{|l|}{\textbf{544}}   & Recall& 0.96& 0.53&0.69 & 0.17&0.23 &0.79    \\ 

\cline{2-8}
\multicolumn{1}{|l|}{} & F1-M& 0.97&0.50 &\textbf{0.49} &0.20 &0.29 &0.73  \\  \hline\hline

\multicolumn{1}{|l|}{\multirow{6}{*}{}}  &Precision &\textbf{1.00} &0.71 &0.36 &0.23 &0.36 & 0.70  \\ 

\cline{2-8}
\multicolumn{1}{|l|}{\textbf{552}}   & Recall& 0.95& 0.63& 0.64& 0.37&0.30 &0.60  \\ 

\cline{2-8}
\multicolumn{1}{|l|}{} & F1-M& 0.97&0.67 &0.46 &0.29 &0.33 &0.64  \\  \hline\hline

\multicolumn{1}{|l|}{\multirow{6}{*}{}}  &Precision  &\textbf{1.00} &0.60 &0.30 & 0.27&0.32& 0.66 \\ 

\cline{2-8} 
\multicolumn{1}{|l|}{\textbf{559}}   &Recall&0.99 & 0.65& 0.70&0.32 &0.21 &0.60  \\ 

\cline{2-8} 
\multicolumn{1}{|l|}{} & F1-M & \textbf{1.00}&0.62 &0.42 &0.29 &0.25 &0.62  \\  \hline\hline

\multicolumn{1}{|l|}{\multirow{6}{*}{}}  & Precision&\textbf{1.00} & 0.82&0.36 &0.36 &\textbf{0.59} &0.81   \\ 

\cline{2-8} 
\multicolumn{1}{|l|}{\textbf{563}}   & Recall& \textbf{1.00}& 0.74& \textbf{0.74}& \textbf{0.44}& 0.34& \textbf{0.84}   \\ 

\cline{2-8}
\multicolumn{1}{|l|}{} & F1-M &\textbf{1.00} &0.78 &\textbf{0.49} &\textbf{0.40} &0.43 &\textbf{0.83}  \\  \hline\hline

\multicolumn{1}{|l|}{\multirow{6}{*}{}}  & Precision &\textbf{1.00} &0.52 &0.34 &0.33 &0.38 &0.80   \\ 

\cline{2-8}
\multicolumn{1}{|l|}{\textbf{567}}   & Recall& \textbf{1.00}& 0.81& 0.60&0.37 &0.26 &0.77   \\ 

\cline{2-8}
\multicolumn{1}{|l|}{} & F1-M &\textbf{1.00} & 0.63& 0.43& 0.35& 0.31& 0.79  \\  \hline\hline

\multicolumn{1}{|l|}{\multirow{6}{*}{}}  & Precision  &0.98 &0.54 &0.20 &\textbf{0.37} &0.36 & \textbf{0.95} \\ 

\cline{2-8}
\multicolumn{1}{|l|}{\textbf{570}}   & Recall& 0.98&0.69 &0.29 &0.41 &0.31 &0.63  \\ 

\cline{2-8} 
\multicolumn{1}{|l|}{} & F1-M & 0.98& 0.61& 0.23& 0.39& 0.33& 0.76  \\  \hline\hline

\multicolumn{1}{|l|}{\multirow{6}{*}{}}  & Precision & 0.99&0.69 &0.34 &0.36 &0.47 &0.75  \\ 

\cline{2-8}
\multicolumn{1}{|l|}{\textbf{575}}   &Recall&0.99 & 0.68& 0.56& 0.39&\textbf{0.41} &0.72   \\ 

\cline{2-8} 
\multicolumn{1}{|l|}{} & F1-M & 0.99& 0.69&0.43 &0.37 &\textbf{0.44} &0.73  \\  \hline\hline

\multicolumn{1}{|l|}{\multirow{6}{*}{}}  & Precision & \textbf{1.00}& \textbf{1.00}& 0.32& 0.22&0.25 &0.59  \\ 

\cline{2-8}
\multicolumn{1}{|l|}{\textbf{584}}   & Recall & \textbf{1.00}& \textbf{0.83}& 0.40& 0.20&0.23 &0.62   \\ 

\cline{2-8}
\multicolumn{1}{|l|}{} & F1-M & \textbf{1.00}&\textbf{0.91} &0.35 &0.21 &0.24 &0.61   \\  \hline\hline

\multicolumn{1}{|l|}{\multirow{6}{*}{}}  & Precision &\textbf{1.00} & 0.34& 0.18& 0.27&0.36 &0.63   \\ 

\cline{2-8}
\multicolumn{1}{|l|}{\textbf{588}}   & Recall & \textbf{1.00}& 0.53& 0.41&0.25 &0.24 &0.56  \\ 

\cline{2-8}
\multicolumn{1}{|l|}{} & F1-M & \textbf{1.00}& 0.42& 0.25& 0.26& 0.29&0.59  \\  \hline\hline

\multicolumn{1}{|l|}{\multirow{6}{*}{}}  & Precision & \textbf{1.00}& 0.66&0.26 &0.29 &0.30 &0.70  \\ 

\cline{2-8}
\multicolumn{1}{|l|}{\textbf{591}}   & Recall & 0.97&0.77 &0.49 &0.32 &0.21 &0.69   \\ 

\cline{2-8}
\multicolumn{1}{|l|}{} & F1-M& 0.99& 0.71& 0.34& 0.31& 0.25&0.69 \\  \hline\hline

\multicolumn{1}{|l|}{\multirow{6}{*}{}}  & Precision &\textbf{1.00} &0.53 &0.29 &0.34 &0.28 &0.70  \\ 

\cline{2-8}
\multicolumn{1}{|l|}{\textbf{596}}   & Recall & 0.99& 0.76&0.57 &\textbf{0.44} &0.20 &0.59   \\ 

\cline{2-8}
\multicolumn{1}{|l|}{} & F1-M &0.99 &0.62 &0.38 &0.38 &0.24 &0.64  \\  \hline

\end{tabular}}
\end{table}

\begin{table}
\centering
\caption[Population-Based LSTM Results with Less Test Data for Each Subject Using 6 Classes]{Population-Based LSTM Results with Less Test Data for Each Subject Using 6 Classes \label{tab:LSTMPop6less}}
\setlength\extrarowheight{-2pt}
\resizebox{!}{7cm}
{\begin{tabular}
{|p{1.5cm}|p{2cm}|p{1.8cm}|p{1.8cm}|p{1.8cm}|p{1.8cm}|p{1.8cm}|p{1.8cm}|}
\cline{1-8}\textbf{Subject} & \textbf{Metric} & 
\multicolumn{6}{ l |} {\textbf{Class}}\\ 
\cline{3-8} & & 0 & 1  & 2 & 3 & 4 & 5   \\ \hline
    
\multicolumn{1}{|l|}{\multirow{3}{*}{}}  & Precision & \textbf{1.00} & 0.54   &0.31  & 0.30 &  0.40& 0.66 \\ 

\cline{2-8}
\multicolumn{1}{|l|}{\textbf{540}}   & Recall & \textbf{1.00}& 0.79& 0.54& 0.33& \textbf{0.34}&0.55   \\ 

\cline{2-8}
\multicolumn{1}{|l|}{} & F1-M &\textbf{1.00} &0.64&0.39 & 0.32& 0.37&0.60  \\  \hline\hline

\multicolumn{1}{|l|}{\multirow{6}{*}{}}  & Precision  & 0.98& 0.48&0.35&0.25 & 0.38& 0.68 \\ 

\cline{2-8}
\multicolumn{1}{|l|}{\textbf{544}}   & Recall& 0.96& 0.53&0.62 & 0.17&0.23 &0.79    \\ 

\cline{2-8}
\multicolumn{1}{|l|}{} & F1-M& 0.97&0.50 &0.44 &0.20 &0.29 &0.73  \\  \hline\hline

\multicolumn{1}{|l|}{\multirow{6}{*}{}}  &Precision &\textbf{1.00} &0.71 &0.33 &0.27 &0.33 & 0.70  \\ 

\cline{2-8}
\multicolumn{1}{|l|}{\textbf{552}}   & Recall& 0.99& 0.71& 0.69& 0.35&0.27 &0.60  \\ 

\cline{2-8}
\multicolumn{1}{|l|}{} & F1-M& 0.99&0.71 &0.45 &0.30 &0.30 &0.65  \\  \hline\hline

\multicolumn{1}{|l|}{\multirow{6}{*}{}}  &Precision  &\textbf{1.00} &0.62 &0.27 & 0.26&0.35& 0.67 \\ 

\cline{2-8} 
\multicolumn{1}{|l|}{\textbf{559}}   &Recall&\textbf{1.00} & 0.70& 0.70&0.32 &0.21 &0.58  \\ 

\cline{2-8} 
\multicolumn{1}{|l|}{} & F1-M & \textbf{1.00}&0.66 &0.39 &0.28 &0.26 &0.62  \\  \hline\hline

\multicolumn{1}{|l|}{\multirow{6}{*}{}}  & Precision&\textbf{1.00} & 0.79&\textbf{0.37} &0.35 &\textbf{0.58} &0.80   \\ 

\cline{2-8} 
\multicolumn{1}{|l|}{\textbf{563}}   & Recall& 0.99& 0.74& \textbf{0.74}& 0.44& 0.33& \textbf{0.84}   \\ 

\cline{2-8}
\multicolumn{1}{|l|}{} & F1-M &0.99 &0.77 &\textbf{0.49} &0.39 &\textbf{0.42} &\textbf{0.82}  \\  \hline\hline

\multicolumn{1}{|l|}{\multirow{6}{*}{}}  & Precision &\textbf{1.00} &0.55 &\textbf{0.37} &0.32 &0.42 &0.79   \\ 

\cline{2-8}
\multicolumn{1}{|l|}{\textbf{567}}   & Recall& \textbf{1.00}& 0.77& 0.67&0.30 &0.32 &0.76   \\ 

\cline{2-8}
\multicolumn{1}{|l|}{} & F1-M &\textbf{1.00} & 0.64& 0.48& 0.31& 0.36& 0.78  \\  \hline\hline

\multicolumn{1}{|l|}{\multirow{6}{*}{}}  & Precision  &0.98 &0.52 &0.20 &\textbf{0.37} &0.35 & \textbf{0.93} \\ 

\cline{2-8}
\multicolumn{1}{|l|}{\textbf{570}}   & Recall& 0.95&0.69 &0.29 &0.41 &0.29 &0.63  \\ 

\cline{2-8} 
\multicolumn{1}{|l|}{} & F1-M & 0.97& 0.60& 0.23& 0.39& 0.32& 0.75  \\  \hline\hline

\multicolumn{1}{|l|}{\multirow{6}{*}{}}  & Precision & 0.99&0.69 &0.36 &0.35 &0.43 &0.74  \\ 

\cline{2-8}
\multicolumn{1}{|l|}{\textbf{575}}   &Recall&0.99 & 0.71& 0.61& 0.35&0.32 &0.75   \\ 

\cline{2-8} 
\multicolumn{1}{|l|}{} & F1-M & 0.99& 0.70&0.46 &0.35 &0.37 &0.74  \\  \hline\hline

\multicolumn{1}{|l|}{\multirow{6}{*}{}}  & Precision & \textbf{1.00}& \textbf{1.00}& 0.32& 0.22&0.25 &0.59  \\ 

\cline{2-8}
\multicolumn{1}{|l|}{\textbf{584}}   & Recall & \textbf{1.00}& \textbf{0.83}& 0.40& 0.20&0.23 &0.62   \\ 

\cline{2-8}
\multicolumn{1}{|l|}{} & F1-M & \textbf{1.00}&\textbf{0.91} &0.35 &0.21 &0.24 &0.61   \\  \hline\hline

\multicolumn{1}{|l|}{\multirow{6}{*}{}}  & Precision &\textbf{1.00} & 0.37& 0.14& 0.24&0.35 &0.64   \\ 

\cline{2-8}
\multicolumn{1}{|l|}{\textbf{588}}   & Recall & \textbf{1.00}& 0.53& 0.29&0.25 &0.26 &0.55  \\ 

\cline{2-8}
\multicolumn{1}{|l|}{} & F1-M & \textbf{1.00}& 0.43& 0.19& 0.25& 0.30&0.59  \\  \hline\hline

\multicolumn{1}{|l|}{\multirow{6}{*}{}}  & Precision & \textbf{1.00}& 0.64&0.29 &0.30 &0.31 &0.70  \\ 

\cline{2-8}
\multicolumn{1}{|l|}{\textbf{591}}   & Recall & 0.99&0.77 &0.45 &0.35 &0.24 &0.68   \\ 

\cline{2-8}
\multicolumn{1}{|l|}{} & F1-M& \textbf{1.00}& 0.70& 0.35& 0.33& 0.27&0.69 \\  \hline\hline

\multicolumn{1}{|l|}{\multirow{6}{*}{}}  & Precision &\textbf{1.00} &0.51 &0.31 &0.36 &0.28 &0.69  \\ 

\cline{2-8}
\multicolumn{1}{|l|}{\textbf{596}}   & Recall & 0.95& 0.73&0.60 &\textbf{0.48} &0.20 &0.59   \\ 

\cline{2-8}
\multicolumn{1}{|l|}{} & F1-M &0.98 &0.60 &0.41 &\textbf{0.41} &0.23 &0.64  \\  \hline
\end{tabular}}
\end{table}

\begin{table}[!ht]
\centering
\caption[Comparison of the Population-Based and Subject-Specific Approach for Each Model Using 6 Classes]{Comparison of the Population-Based and Subject-Specific Approach for Each Model Using 6 Classes \label{tab:Comp6classless}}
\setlength\extrarowheight{-2pt}
\begin{tabular}{p{0.5cm}|p{1.6cm}|p{0.9cm}|p{0.9cm}|p{0.9cm}|p{0.9cm}|p{0.9cm}|p{0.9cm}|p{0.9cm}|}
\cline{2-9}
    & \textbf{Metric} & \textbf{Avg} &
    \multicolumn{6}{ l |} {\textbf{Class}}\\ 
    \cline{4-9} & & & 0& 1  & 2 & 3 & 4 & 5 \\ \hline

\multicolumn{1}{|l|}{\multirow{4}{*}{}} & Precision & 0.51  & 0.94  & 0.47 &0.30&0.34 &0.35  &0.66  \\  

\cline{2-9} 
\multicolumn{1}{|l|}{\textbf{PB ResNet}} & Recall  &  0.54 &  0.96 & 0.58 & 0.41&0.33& 0.22 & 0.72  \\ 

\cline{2-9}
\multicolumn{1}{|l|}{}  & F1-M  & 0.52 &  0.95& 0.52  & 0.35 & 0.33 & 0.27 & 0.69  \\   
 \hline\hline

\multicolumn{1}{|l|}{\multirow{4}{*}{}} &  Precision & 0.49  & 0.94  & 0.39 &0.27&0.30 &0.37  &0.68 \\    

\cline{2-9}  
\multicolumn{1}{|l|}{\textbf{SS ResNet}} & Recall  &  0.53 &  0.90 & 0.62 & 0.42&0.32& 0.25 & 0.69   \\

\cline{2-9} 
\multicolumn{1}{|l|}{}  & F1-M  & 0.50 &  0.92& 0.48  & 0.33 & 0.31 & 0.30 & 0.69   \\
\hline\hline

\multicolumn{1}{|l|}{\multirow{4}{*}{}} & Precision  & 0.55& 1.00 & 0.60  & 0.31 & 0.31 & 0.37 & 0.71\\ 

\cline{2-9} 
\multicolumn{1}{|l|}{\textbf{PB LSTM}}   & Recall  & 0.59  &0.99& 0.73  & 0.56 &0.34 & 0.28 & 0.66  \\  

\cline{2-9} 
\multicolumn{1}{|l|} {} & F1-M  &  0.56 & 0.99  & 0.66 & 0.40& 0.32 &0.32& 0.68 \\  \hline\hline

\multicolumn{1}{|l|}{\multirow{4}{*}{}} & Precision  & 0.55& 1.00 & 0.60  & 0.30 & 0.31&0.38  &0.71  \\ 

\cline{2-9}
\multicolumn{1}{|l|}{\textbf{SS LSTM}}    & Recall  &   0.59&0.99&  0.71 & 0.56 & 0.35& 0.29 & 0.66  \\  

\cline{2-9}
\multicolumn{1}{|l|}{} & F1-M  &  0.56 & 0.99  & 0.65 & 0.39& 0.32 &0.33& 0.68 \\    \hline\hline

\multicolumn{1}{|l|}{\multirow{4}{*}{}} & Precision  &  0.52 & 0.97  & 0.46 & 0.30& 0.32 & 0.37 &0.69  \\

\cline{2-9}
\multicolumn{1}{|l|}{\textbf{PB Hybrid}}   & Recall  &  0.56 & 0.95  & 0.70  & 0.44 & 0.34 & 0.27&0.68  \\  

\cline{2-9} 
\multicolumn{1}{|l|}{} & F1-M  &  0.53 &  0.96 & 0.55& 0.36& 0.33& 0.31 & 0.69 \\     \hline\hline

\multicolumn{1}{|l|}{\multirow{4}{*}{}} & Precision  & 0.50  & 0.96  &0.39  & 0.28&  0.30& 0.38 &0.70  \\ 

\cline{2-9}  
\multicolumn{1}{|l|}{\textbf{SS Hybrid}}    & Recall  &0.54   &  0.88 & 0.63 & 0.46 &0.35 &0.27 &0.65   \\  

\cline{2-9}
\multicolumn{1}{|l|}{} & F1-M  & 0.51  & 0.92  & 0.49& 0.35& 0.33& 0.32 &0.68 \\  \hline
\end{tabular}
\end{table}

\newpage
\section{Discussion}
\label{chap:Discussion}

Hypoglycemia prediction is a major challenge in diabetes research, with significant variations among the subjects.
\\
One limitation of ML models is that data is not AI-ready and requires multiple preprocessing steps when using sensor data. Notably, the OhioT1DM dataset collected data from two different cohorts employing different activity trackers. Therefore, exercise was measured with different metrics and could not be fused without converting step count into acceleration. Still, variations between both parameters existed, necessitating more extensive preprocessing. 
Moreover, large gaps in glucose data result in information loss and prevent using all data-frames. To compensate for the gaps, linear interpolation was applied. However, better methods should be used since glucose does not behave linearly. 
\\
Moving to the DL models, the training process revealed notable discrepancies between training, validation, and test performance. Overfitting was evident when trained for longer epochs, necessitating early stopping and reducing training time. Some subjects were trained for fewer than five epochs, resulting in suboptimal performance. Training accuracy ranged from 30–40\% with nine classes and 60–65\% with six classes. However, validation loss did not consistently reflect training loss, suggesting high inter-subject variability in hypoglycemic events. Even subject-specific models could not generalize within validation data, likely due to small sample sizes, class imbalance, and limitations of DL models. Possibly, the validation data did not represent the learned patterns of the training data. Popular subjects could lead to a bias so that less represented samples are not learned with strong connections. Class weights affected bias, as models without weighting classified most instances into the dominant class. 
Hence, in the population-based approach, the patterns leading to the onset of hypoglycemia in short-term PHs are similar in most of the subjects and the prediction 0-30 before onset is feasible. In contrast, it is more challenging to develop a model that is capable of predicting multiple hours before the event, which may need a larger dataset and individual training. 
\\
Population-based models identified hypoglycemia onset within 5–30 minutes with high recall but lower precision, indicating frequent false alarms. While short-term predictions benefited from subject-specific models, long-term forecasting remained challenging. Notably, subject 570 showed the greatest improvement despite a small dataset with only 227 hypoglycemic data points, suggesting that the representativeness of the data samples is influential. In addition, subjects 540, 575, and 591, having the most representative datasets, improved significantly as well, indicating that larger datasets remain advantageous. Conversely, models trained with six classes exhibited minimal impact from individualized training, with slight improvements in subjects 570, 563, and 596. Subject-specific models mainly benefit the classification of class 4, looking at the average performance across subjects. 
Individualization introduced limitations, as subjects with fewer samples trained for fewer epochs, potentially diminishing their influence. Removing dominant classes in the 6-class approach further reduced sample availability, limiting training efficacy. Individual models trained only on test subject data might yield better results. However, current sample sizes are insufficient for robust training.
\\
The macro averages for the population-based model using nine classes indicated suboptimal performance and bias induced by class imbalance. Even dominant classes were poorly classified, suggesting insufficient class distinction. The population-based LSTM model classified 50–68\% of hypoglycemic events 15–30 minutes prior, with individualized models improving to 50–72\%. With six classes, prediction rates increased to 56–73\% for population-based models and 56–71\% for subject-specific models. The macro average recall of all classes improved from 20\% to 59\% using fewer classes, with up to 70\% of events predicted within 15 minutes, while nearly 80\% were detected in subject 552. Most subjects reached 60–70\% for 30-minute predictions.
\\
It was also noticed that the performance among DL models across subjects varied. Consequently, individually tuned and selected models for each subject are suggested. The LSTM model slightly outperformed the ResNet model with nine classes and significantly outperformed it with six classes. Subject 567 achieved the best overall results. For population-based models, subject 563 performed best with the ResNet model, while subject 567 with the LSTM model. Individualized training improved the performance of the LSTM model for subject 570 and the ResNet model for subject 575. With six classes and subject-specific approaches, subjects 563 and 567 performed best with LSTM, while ResNet and hybrid models remained superior for subject 575.
Examining the best-performing subjects, subject 567 has the third-highest hypoglycemic event count but also significant missing glucose data. Fewer instances of class 1 are available which could indicate that most events were of longer duration and could be severe hypoglycemia. Similarly, subject 575 has the highest hypoglycemic event count and fewer class 1 samples. These observations suggest that model performance improved when excluding dominant subjects and that highly represented subjects may introduce biases that hinder population-based classification.
\\
Challenges remain, and precision is still low, even using fewer classes. However, it needs to be considered that the model was not tuned for training six classes. Lastly, the dataset size is insufficient to teach all observed patterns and behaviors. Nevertheless, the results are promising and propose that a classification of hypoglycemia onset up to 4h before is feasible with better data and model architectures. 

\newpage
\section{Conclusion and Future Work}
\label{chap:Conlusion}

This study developed a DL system to classify hypoglycemia onset, incorporating nine classes ranging from 0 to 24 hours before an event. Thus, the model supports both short-term preventive actions and long-term planning for meals and activities. The model was trained on 12 subjects from the OhioT1DM dataset using glucose, insulin, and activity data. LSTM, ResNet, and a hybrid model were compared for population-based and subject-specific approaches.
Results showed that LSTM models generally outperformed others, with better performance in individualized models. Subjects with more data achieved higher accuracy with subject-specific models since those introduced a bias for population-based models. Predictions were more accurate for short-term horizons (0–30 min before onset), whereas long-term classifications led to misclassifications with the proposed methods and data. Classifying only up to 4 hours before hypoglycemia onset, the precision increased as well as the overall performance. Most errors were time-shifted, meaning predictions were often within the nearest neighbors. Comparing individual models, performance did not change significantly, while class 4 mainly profited from personalization. Short-term models predicted 60\% of hypoglycemic events, but integrating short- and long-term horizons in a single model resulted in insufficient performance. Thus, the short-term model should be tuned with better datasets and model architectures.
\\
Future work should explore separate models for short- (0–2h) and long-term (4–24h) predictions. A layered classification system could classify between short-, long-term, or no risk before applying specialized models predicting either the short- or the long-term onset of hypoglycemia. Additionally, conventional ML models should be explored since DL tends to overfit small datasets.
Further improvements could involve better data imputation methods for missing glucose values, given the non-linear nature of glucose patterns. A possible method could be a cubic interpolation, linear regression for time series data, or random forest regression predicting the possible values while considering the time.

\bibliographystyle{unsrt}  
\bibliography{references}

\end{document}